\documentclass[%
 aip,
 amsmath,amssymb,
 reprint,%
]{revtex4-1}

\usepackage{graphicx}
\usepackage{dcolumn}
\usepackage{bm}

\usepackage[utf8]{inputenc}
\usepackage[T1]{fontenc}
\usepackage{mathptmx}
\usepackage{etoolbox}
\usepackage{hyperref}
\usepackage{tikz}
\makeatletter
\def\@email#1#2{%
 \endgroup
 \patchcmd{\titleblock@produce}
  {\frontmatter@RRAPformat}
  {\frontmatter@RRAPformat{\produce@RRAP{*#1\href{mailto:#2}{#2}}}\frontmatter@RRAPformat}
  {}{}
}%
\makeatother
\begin{document}

\preprint{AIP/123-QED}

\title[Friction in folded graphene]{Measuring friction from simulations of folded graphene sheets}
\author{Charlie M. Rawlins}
\author{Gareth A. Tribello}%
 \email{c.rawlins@qub.ac.uk}
\affiliation{ 
Centre for Quantum Materials and Technologies, School of Mathematics and Physics, Queen's University Belfast, Belfast, BT7 1NN 
}%

\date{\today}

\begin{abstract}
We run molecular dynamics simulations of folded graphene sheets and present a procedure to measure the sliding friction in these systems based on the rate of decay of a damped-harmonic oscillator. 
This procedure allowed us to study the affect the size, geometry and the temperature of the graphene sheet had on the ability to propagate the initial fold and the rate at which it settles to a final 'fully-folded' equilibrium state.  We offer simple rationalisations for the relationships between the initial geometries of our simulations and the friction values that emerge. 
\end{abstract}

\maketitle

\section{\label{sec:intro}Introduction}

The discovery of graphene has led to a flurry of new activity in physics, chemistry and materials research\cite{Yanwu_10}. This novel material has numerous unique properties many of which come about because a graphene sheet is only one atom thick\cite{Neto_09}.  This 2D geometry ensures that graphene has excellent in-plane mechanical flexibility, which, because  there are also strong inter-layer van der Waals interactions between graphene sheets, ensures graphene sheets can be scrolled\cite{Viculis_03, Pereira_21}, folded\cite{Zhang_10} and stacked into novel assemblies.  These assemblies of graphene also have interesting properties one of the most intriguing of which is the evidence of super lubricity\cite{Dienwiebel_lubricity_04,Yilun_14}. 

Recently Annett and Cross have developed a method for synthesising folded graphene\cite{Cross_16}. They first disrupt the graphene structure using a nanoindentor or AFM tip.  Novel kirigami ribbons then grow spontaneously from the disrupted region. Annett and Cross have used continuum mechanics to rationalise the geometry of the structures that form. However, this model provides no information on the atomic scale features that affect the formation and growth of these ribbons.  Atomistic simulation thus has a clear role to play when it comes to better understanding this phenomenon.

Simulating the formation of kirigami ribbons is difficult because when the carbon-carbon bonds break carbon likely forms bonds with gaseous species from the atmosphere.  Incorporating these bond formation events in simulations is extremely difficult, which is likely why the results from early simulations are mixed\cite{Fonseca2018SelftearingAS, Fonseca_18, He_18}.  However, even though it is likely not possible to simulate kirigami ribbon formation directly, carefully performed simulation can still provide insight as it allows one to study geometries that are simpler than those generated by experiment.  
For example, in the case of multilayer graphene, simulations have been performed where the top layer of graphene has been laterally offset.
The resulting simulations show that the top layer will retract to be in alignment with the lower layers as the aligned configuration is more energetically favourable\cite{Yang_13_lubricity,Yang_12_lubricity, Zheng_08_self_retract, Ng_12_Self_retract}.  Furthrmore, as it returns to this equilibrium state the graphene sheet will undergo oscillatory motion \cite{Lebedeva_11}. 
The dynamics of the motion has been related to factors such as commensurability \cite{Xu_13_commensurate,Popov_11_commensurate,Popov_11_commensurateb}, temperature\cite{Yang_13_lubricity} and surface roughness\cite{Ye_2014, Ye_2015}.
This work has helped to provide insight into the interlayer forces of graphene, however, in the case of ribbon-formation, the fold plays an important role in determining the directionality and motion of the sheet. This fold is noticeably absent from these simulations and experiments.

Studies have been done on the self-folding and self-scrolling of graphene sheets when an edge has been deformed \cite{Pereira_21}. However, the motion of these kinds of folded and scrolled sheets has not been studied in great detail. Only the initial and final structures of edge-deformed graphene sheets have been investigated.  The intermediate motions which govern the rate at which these structures form has not been explored.

In this work, we study the behaviour of a folded sheet of graphene in vacuum. We find that these sheets exhibit similar motion to free-standing sheets and undergo damped harmonic oscillation as they reach an energetically more stable configuration. 
The model of this motion can be described using parameters which depend on the graphene twist angle, the dimensions of the sheet and the temperature. The parameters used for describing the dampening of the system can be used to provide information on the amount of friction between the layers.

In what follows we show that friction is lowered when graphene sheets are non-commensurately stacked and when temperature is reduced.  The remainder of this paper is laid out as follows.  We first describe how the initial graphene sheet is set-up and how the resulting structures are characterised in section \ref{sec:config}. Secondly, we describe the procedure used to extract useful data from the simulations and how we model the data using a damped-harmonic oscillator in section \ref{Sec:fric}. 
In the final parts of the paper we then examine how different initial set-ups affect these parameters.

\section{Initial configuration and outcomes}
\label{sec:config}

\begin{figure*}
\includegraphics[scale=0.3]{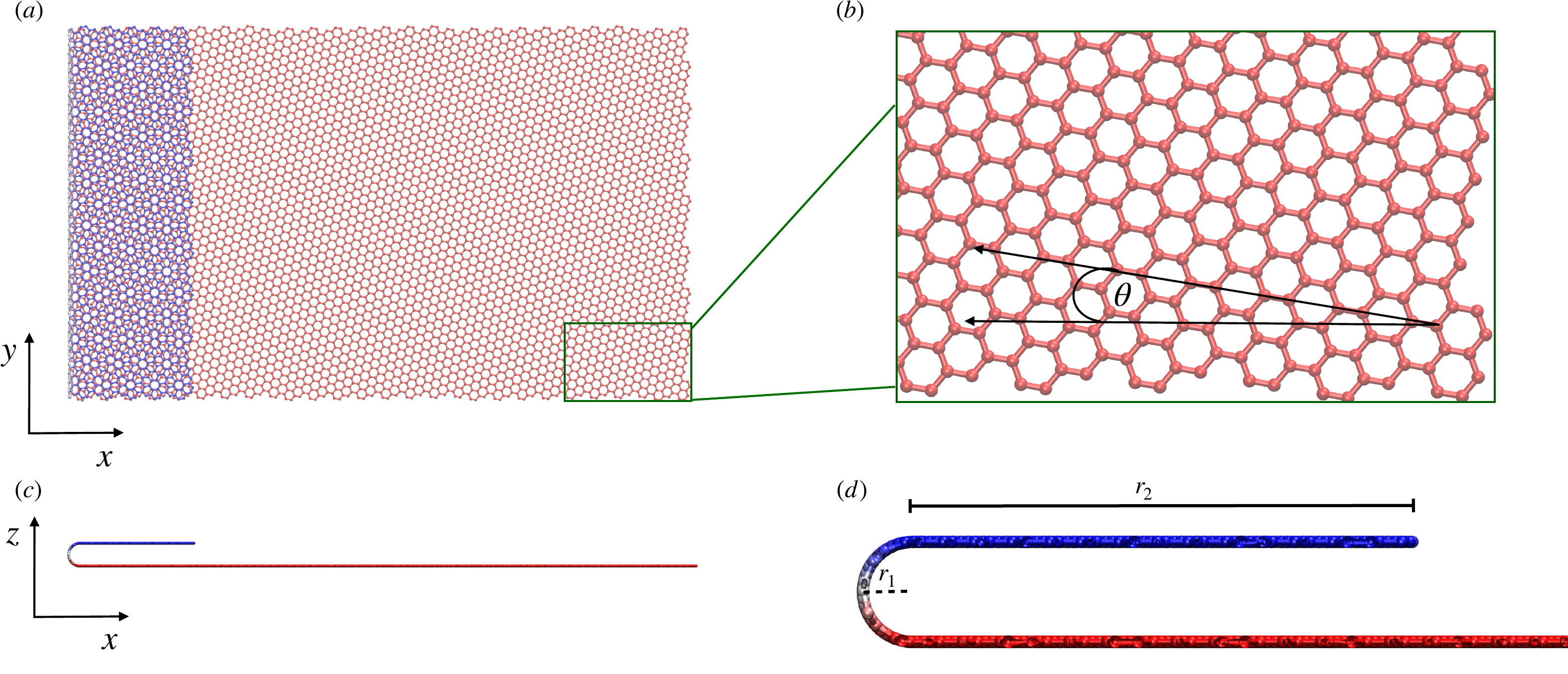}
\caption{\label{fig:layout} Schematic representation of a folded graphene sheet viewed on the $xy$ plane (a) and the $xz$ plane (c). Inserts shows the graphene lattice and how the twist angle $\theta$ (b) and the fold parameters $r_1$ and $r_2$ (d) are defined.}
\end{figure*}


The approach we take in this work is similar to the approach adopted by \citet{Pereira_21}.  We perform MD simulations to investigate how unsupported, single-layer graphene approaches equilibrium from an initial structure.  However, instead of starting the simulation with a configuration in which one end of the sheet is rolled up to form a scroll, we start our simulations from a folded configuration similar to the one shown in figure \ref{fig:layout}.
  
The initial structure is defined by five parameters that describe the geometry of the fold and parameters that describe the shape of the flat sheet.  The first two of these parameters are the length ($l_x$) and width ($l_y$) of the sheet.  A parameter that defines the orientation of the fold relative to the periodic arrangement of carbon atoms is also required. In this work we use the twist angle ($\theta$) for this purpose, which is defined as the angle between the x-axis of the lab frame and the line of symmetry that, if cut, would give rise to an edge with an 'arm-chair' configuration. Cutting graphene sheets along non arm-chair directions leads to structures in which some carbon atoms only participate in only one covalent bond.  We removed these carbons prior to starting simulations and thereby ensured that all the carbon atoms at the edge of the graphene were bound to at least two additional carbon atoms. The fold was then formed parallel to the $y$-axis of the lab frame and defined by two parameters, the radius of curvature $r_1$ and the length of sheet on the top layer which overlaps with the bottom layer $r_2$.  These parameters are illustrated in figure \ref{fig:layout}.

To study the self-folding of graphene sheets, fully-atomistic MD simulations using the ReaxFF potential\cite{ReaxFF_1,ReaxFF_2,Aktulga12}, as implemented in the Large-scale Atomic/Molecular Massively Parallel Simulator (LAMMPS\cite{LAMMPS}), using the parameter set provided by \citet{Chenoweth_08} for C/H/O.  The equations of motion were numerically integrated using the velocity-Verlet integrator and a time-step of 0.125 fs.  Unless otherwise stated temperature was fixed to 300 K using a Nos\'e-Hoover thermostat\cite{Thermostat} with a relaxation time of 125~fs. 

\begin{figure*}
    \includegraphics[scale=0.65]{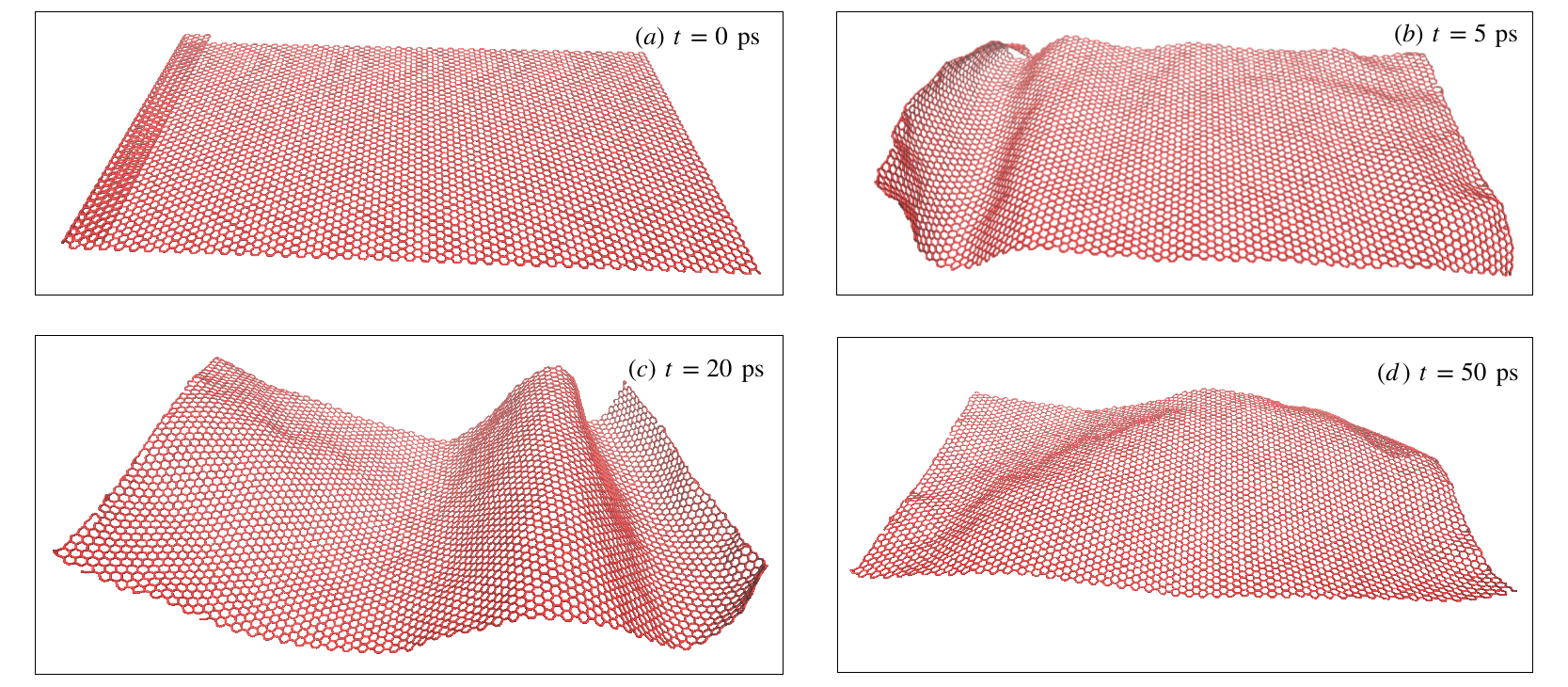}
    \caption{\label{fig:unfurl} VMD snapshots of a 200 \AA\ by 100 \AA\ sheet with initial configuration of $r_1=2.0$, $r_2=10$ and $\theta=0^{\circ}$. This is an example of 'unfolding'.}
\end{figure*}

We ran simulations of a 200 by 100\AA\ graphene sheet starting for twist angles of 0, 10, 20 and 30 degrees and a range of $r_1$ (2 to 3\AA\ ) and $r_2$ (10 to 40\AA\ ) values. For initial testing with $\theta=0$, the range of $r_1$ values was from 1.5 up to 5\AA\ . These simulations were run until the system has equilibrated and the fluctuations in the potential energy were less than 2 meV/atom.  We observed three different behaviours in these simulations.  When the $r_1$ parameter is large and the $r_2$ parameter is small the sheet unfolds in the manner illustrated in figure \ref{fig:unfurl}.  This behaviour makes sense as the overlap between the top and bottom layers is small.  The energetic cost associated with bending the sheet cannot, therefore, be compensated for by the interlayer attraction.

\begin{figure*}
    \includegraphics[scale=0.35]{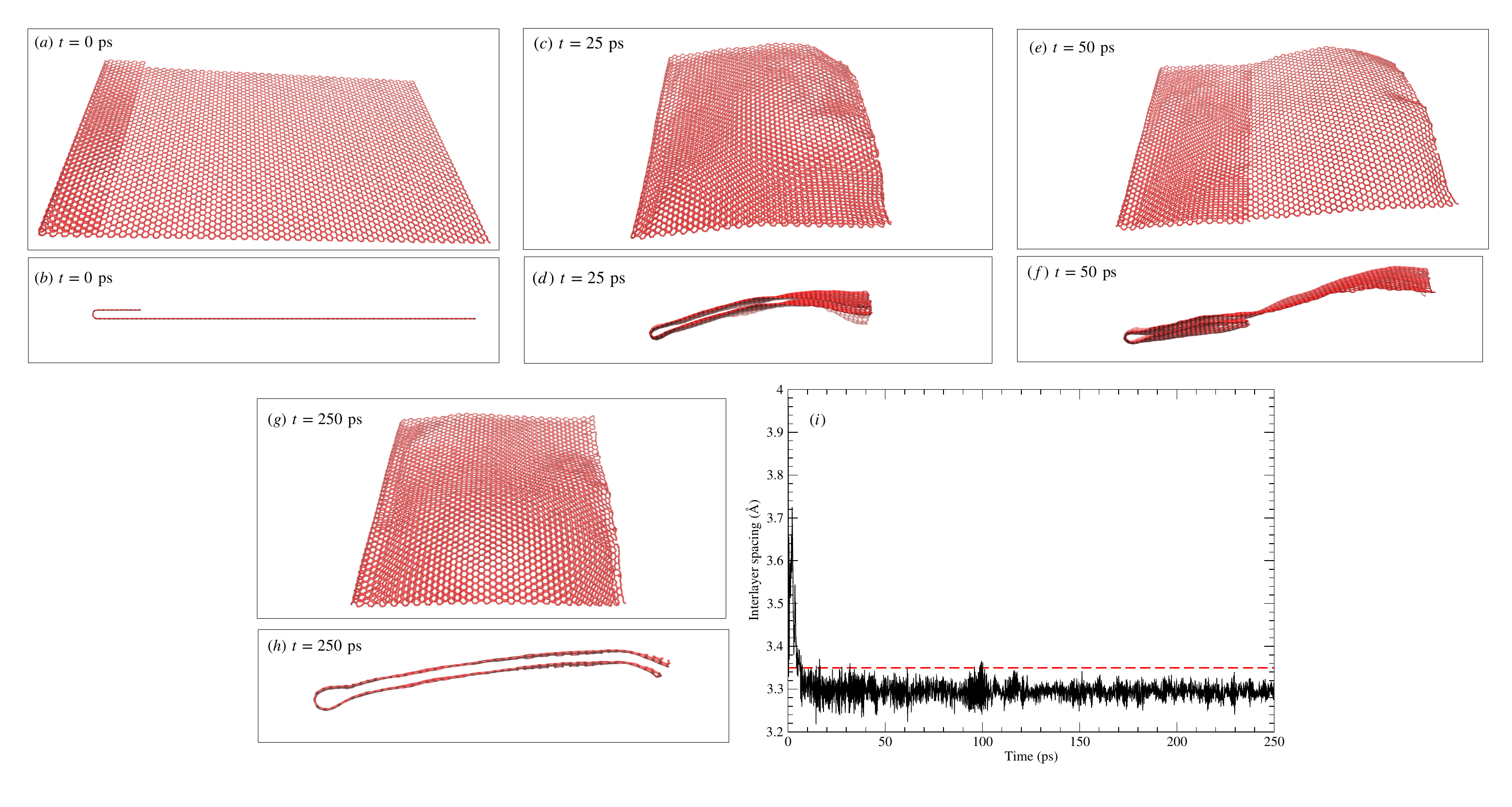}
    \caption{\label{fig:singfold} VMD snapshots of a 200 \AA\ by 100 \AA\ sheet with initial configuration of $r_1=2.0$, $r_2=20$ and $\theta=0^{\circ}$. This is an example of 'folding'. Inset (i) is a plot of the interlayer spacing between the top and bottom layers as a function of time.}
\end{figure*}

When $r_2$ is around 10 times larger than $r_1$, then the top part of the folded sheet slides back and forth along the $x$-axis and over the bottom part of the sheet as illustrated in figure \ref{fig:singfold}.  This same behaviour is consistently observed in simulations started with different initial velocities, which suggests that the energetic cost for folding can be compensated for by an increase in interlayer attraction.  This result is consistent with the results from experiments in which  kirigami structures form.  In those experiments the energy released when the overlap between graphene sheets increases drives carbon-carbon bond breaking reactions.  Consequently, it is hardly surprising that overlap between the top and bottom layers compensates for folding.
Also shown in figure \ref{fig:singfold} are slices along the $xz$ plane for each snapshot. These snapshots show that the spacing between the top and bottom layers is not constant throughout the length of the sheet. Close to the fold the sheet resembles a tennis-racket. However, when this part of the sheet is excluded, we find that the interlayer spacing remains mostly constant throughout the course of the simulation. This is highlighted in figure \ref{fig:singfold}(i), which shows that the interlayer spacing (the method that this is determined is described in the following section)very quickly settles from the initial configuration to a value of about 3.3\AA\ . The expected interlayer spacing of 3.35\AA\ for graphene is shown as a red line in this figure for reference.

\begin{figure*}
    \includegraphics[scale=0.35]{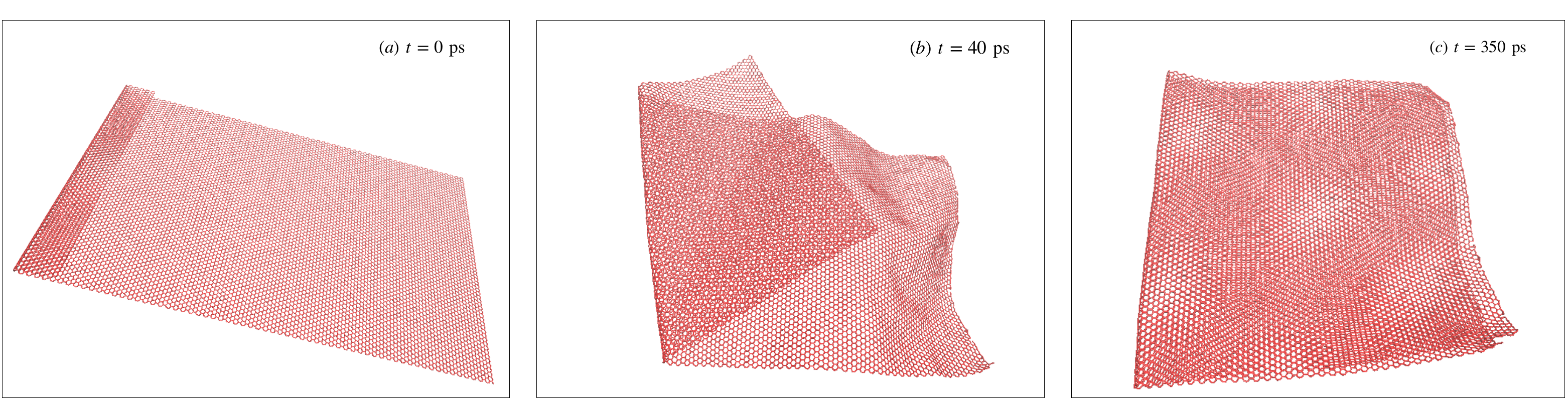}
    \caption{\label{fig:atyp} VMD snapshots of a 300 \AA\ by 150 \AA\ sheet with initial configuration of $r_1=3.0$, $r_2=20$ and $\theta=0^{\circ}$. This is an example of 'atypical folding'.}
\end{figure*}

In the intermediate region, neither regular folding or unfolding is observed. We instead get a behaviour that we have christened 'atypical folding' (see Fig. \ref{fig:atyp}) in which random, in-plane thermal oscillations shift the fold direction away from the $x$-axis.  Analysing these trajectories using the methods that will be explained in the later parts of this paper is difficult which is why we classify these trajectories differently from the ones in the previous paragraph. 

\begin{figure}
    \includegraphics[scale=0.35]{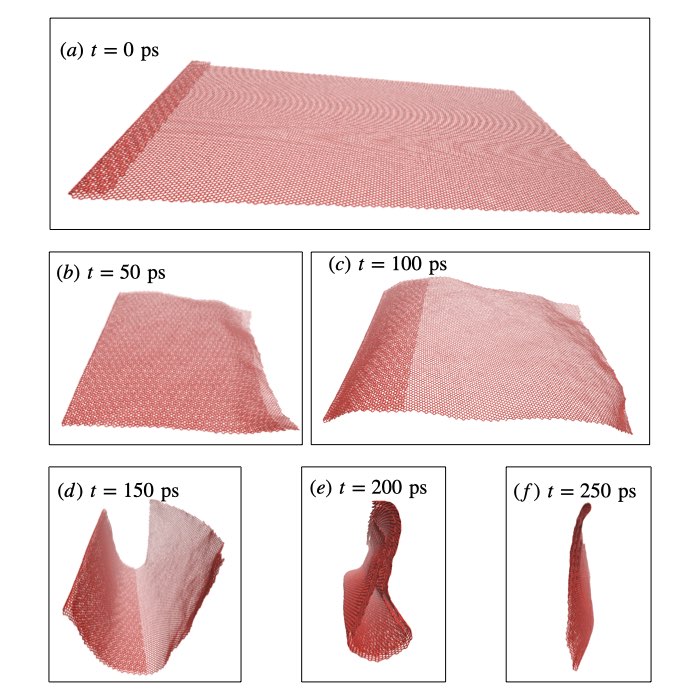}
    \caption{\label{fig:doubfold} VMD snapshots of a 400 \AA\ by 200 \AA\ sheet with initial configuration of $r_1=2.0$, $r_2=20$ and $\theta=10^{\circ}$. The figures show an example where 'double folding' starts after about $t=150 $ ps.}
\end{figure}

No restraints were placed on the carbon atoms in any of our simulations.  The sliding of the graphene layers over each other is thus accompanied by out of plane oscillations when both folding and atypical folding occurs. In fact, for the largest sheets we studied in work (400 \AA\ by 200 \AA\ ) when $\theta=10$ (which we show in later sections to be a low friction set-up) these oscillations are large enough to cause the sheet to fold for a second time as illustrated in figure \ref{fig:doubfold}.     

\begin{figure}
    \includegraphics[scale =0.35]{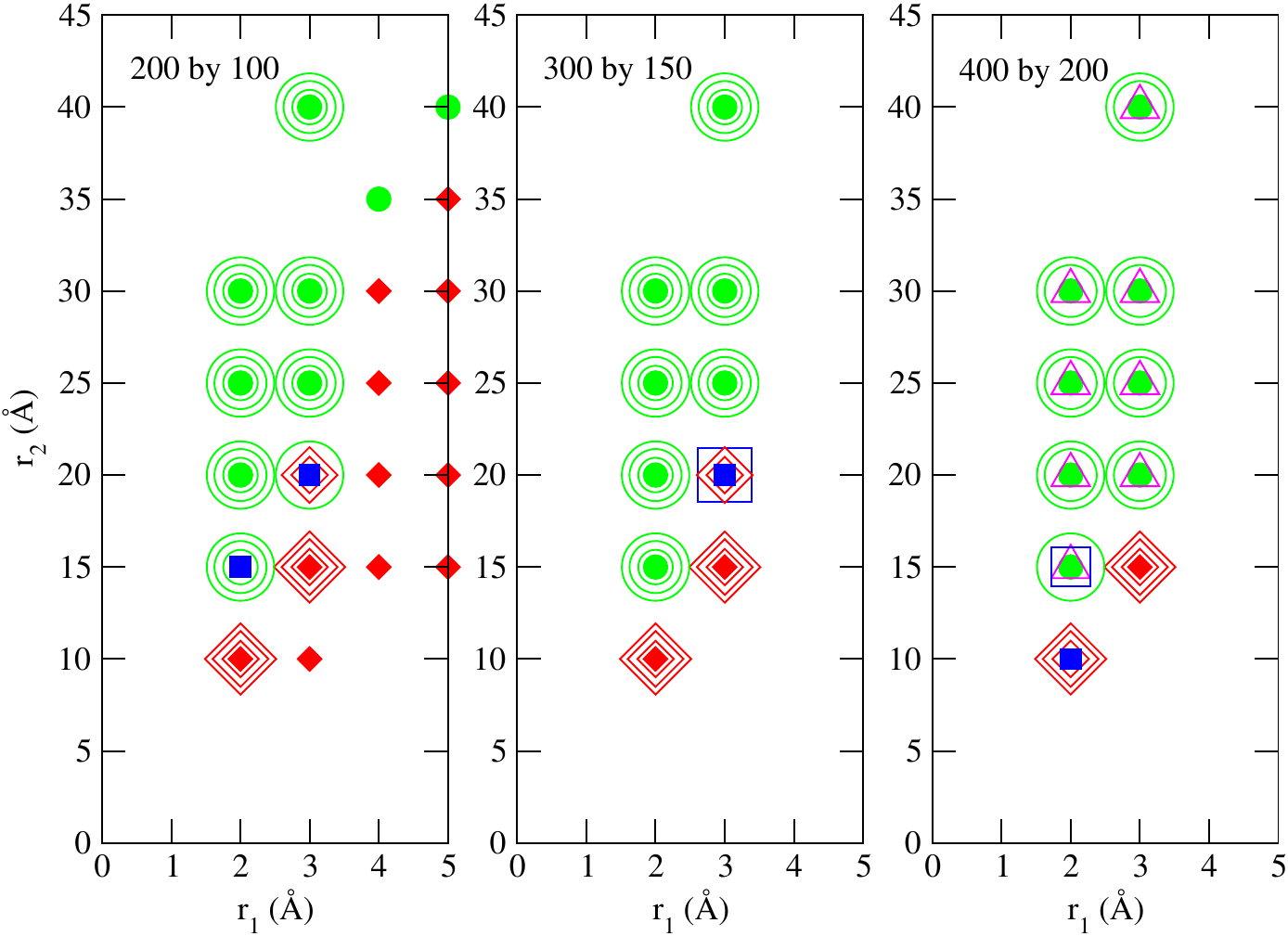}
    \caption{\label{fig:foldcheck} Summary of the final states of an edge-folded graphene sheet based on initial geometries. The calculations we report on here were performed using sheets of dimensions 200 by 100\AA\ (left panel), 300 by 150\AA\ (middle panel) and 400 by 200\AA\ (right panel). Trajectories where folding, unfolding, atypical folding and double folding are represented by green circles, red diamonds, blue squares and magenta triangles respectively. The center, solid symbols are for offset angles of $\theta=0^{\circ}$, the smallest ring around this symbol corresponding to $\theta=10^{\circ}$, the middle sized ring for $\theta=20^{\circ}$ and the largest ring being for $\theta=30^{\circ}$.}
\end{figure}

Figure \ref{fig:foldcheck} summarises the relationships between the initial geometry and the final equilibrium structure the system adopts.  This figure shows that, as discussed in the previous paragraphs, unfolding occurs when $r_2$ is small or $r_1$ is large.  In other words, the folded structure is only retained when there is enough attraction between the graphene layers to compensate for the energetic cost associate with the bending.

 Figure \ref{fig:foldcheck} also includes data from simulations of sheets with larger $l_x$ and $l_y$ values. Specifically 300 by 150 \AA\ and 400 by 200 \AA\ sheets. The results  for these larger sheets are very similar to the results that were obtained for the smaller system.  At first glance this result appears surprising as making the sheet wider increases  interactions between the layers and further stabilises the fold.  However, increasing the width of the sheet also increases the width of the fold.  Having a wider fold increases the energy and, it would appear, counteracts the negative terms that come from having more overlap.   
  
\section{Estimating friction}\label{Sec:fric}

\begin{figure*}
\begin{tikzpicture}
\draw (10,0) -- (0,0) arc (270:90:1) -- (2,2);
\draw[<-|] (10,-0.5) node[anchor=west]{$S_m$} -- (0,-0.5) arc (270:90:1.5) -- (2,2.5) node[anchor=west]{0};
\draw[ultra thick] (10,0) -- (8,0);
\draw[ultra thick] (2,2) -- (0,2);
\draw[dashed] (1.5,2.5) node[anchor=north west]{$S_i$}-- (1.5,-0.5) node[anchor=south west]{$S_j$};
\draw[dashed] (1, 2) -- (1, 1) node[anchor=east]{$r$} -- (1,0); 
\draw[|-|] (2,1.5) -- (10,1.5);
\node[draw=none] at (6,1.65) {$d$};
\draw[|-|] (-2,0) node[anchor=east]{$z_{\rm min}$} -- (-2,2) node[anchor=east]{$z_{\rm max}$};
\draw[|-|] (0,3) -- (1,3) node[anchor=south]{$\sigma$} -- (2,3);
\draw[|-|] (8,0.5) -- (9,0.5) node[anchor=south]{$\sigma$} -- (10,0.5);
\draw[->] (-2.5,1) node[anchor=east]{$x$} -- (-1.8,1);
\end{tikzpicture}
\caption{\label{fig:dist_diag} Schematic to demonstrate how the distance $d$ between the top and bottom edge were determined. Bold lines on the folded sheet correspond to the initial top, flat portion of the sheet and the equivalent portion of graphene on the bottom part of the fold.}
\end{figure*}
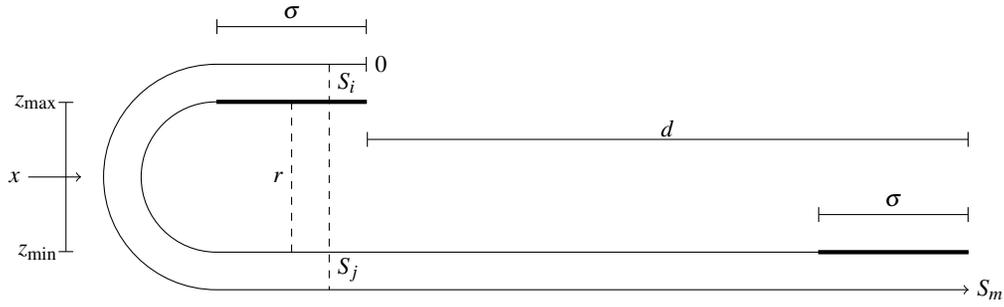

\begin{figure}
    \centering
    \includegraphics[scale=0.35]{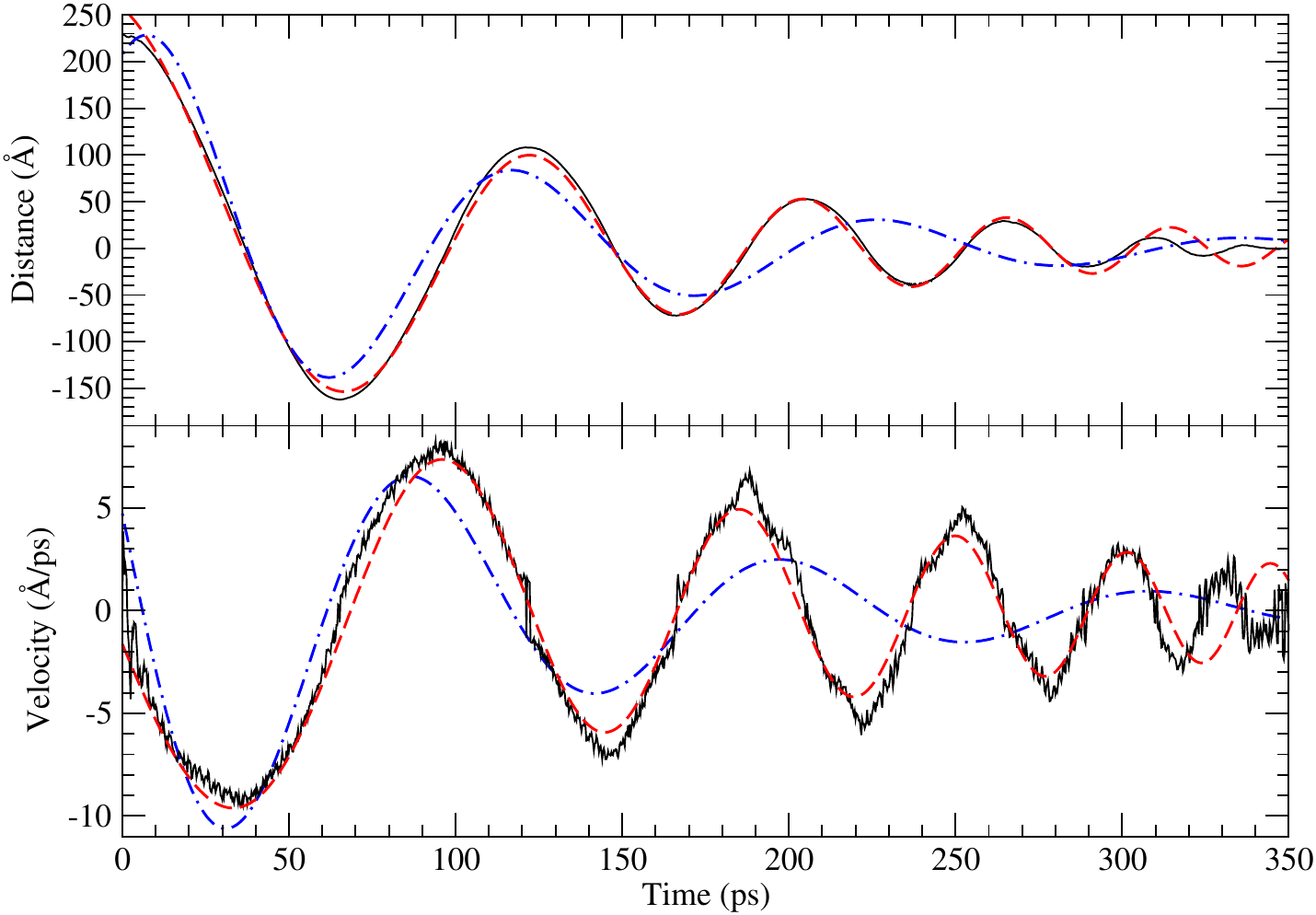}
    \caption{Example of distance between the short edges of the sheet $d$ and sliding velocity over time. The raw data is shown in black, with the red dashed lines being the fit of a damped harmonic oscillator given by Eq. \ref{Eq:xfit} for the top figure and Eq. \ref{Eq:vfit} for the bottom. The blue line is a fit with $\beta=0$.}
    \label{fig:fitexample}
\end{figure}


As shown in figure \ref{fig:singfold}, as the sheet slides over itself the system eventually settles into a fully-folded configuration. 
In order for this to occur, there must be some dampening force to slow down the sliding sheet, which we can assume is due to the friction between the top and bottom layers.
Our contention in the remainder of this paper is that the strength of the dampening force can be determined by examining how the distance between the short edges of the sheet ($d$ in figure \ref{fig:dist_diag}) changes with time.


The method for determining $d$ from our simulations is slightly more involved than figure \ref{fig:dist_diag} suggests.  We are helped by the fact that the atoms in the bold parts of the graphene sheet shown in figure \ref{fig:dist_diag} always remain in the top/bottom layer of the graphene structure as the system oscillates back and forth.  In other words, the atoms in these parts of the sheet can be said to be 'safe' as they never become part of the curved portion of the sheet. These bold parts of the sheet have a length of $\sigma$ and the indices of the atoms within them can be identified from the starting configuration.  This identification is possible because z-coordinates of all the atoms in the initial configurations satisfy $z_{\rm min} \le z \le z_{\rm max}$. Atoms that have $z=z_{\rm min}$ are part of the bottom layer, atoms with $z=z_{\rm max}$ are part of top layer and atoms with $z_{\rm min} < z < z_{\rm max}$ are in the fold.  For the calculation of $d$ we make a list of the indices of the atoms that have $z=z_{\rm max}$ in this initial structure.  We then make a second list that contains the indices of those atoms that have $z=z_{\rm min}$ and that have an $x$ coordinate that is within $\sigma$ of the maximum $x$ coordinate for the atoms.  

We introduce a new coordinate, $S$, for each atom in the graphene sheet.  This coordinate tells us the $x$ position the atom would have if the graphene sheet were laid flat and is, therefore, independent from the shape of the sheet. The $S$ value for each atom is thus determined at the start of the simulation and never changes.  The $S$ coordinates for the atoms in the bottom layer of the initial structure (i.e. those atoms with $z=z_{\rm min}$) can be set equal to the $x$ coordinates of these atoms in the initial structure.  For the remaining atoms, we can measure the distance $dS$ between the initial position of atom $a$, whose $S$ coordinate is known, and one of its neighbouring atoms $b$, whose $S$ coordinate is not known as $dS = \sqrt{(x_a-x_b)^2 + (z_a-z_b)^2}$.  The $S$ coordinate for atom $b$ is then set equal to $S_b = S_a - dS$. This procedure resembles the method that is used to calculate geodesic distances in the ISOMAP algorithm \cite{isomap}.

When calculating the quantities labelled $d$ and $r$ from the instantaneous coordinates of the atoms we use the indices of the 'safe' atoms in the blue regions of figure \ref{fig:dist_diag} and the $S$ coordinates that were determined from the initial structure.  To calculate $r$ we first calculate an $N_s \times N$ matrix of distances, $D$, where $N_s$ is the number of 'safe' atoms and $N$ is the total number of atoms in the structure.  The distances in this matrix $\mathbf{D}$ are calculated using Pythagoras' theorem and the instantaneous coordinates of the atoms.  We then calculate a second $N_s \times N$ matrix $\mathbf{M}$ as:
$$
M_{ij} = \begin{cases}
D_{ij} & \textrm{if} \qquad | S_i - S_j | > \frac{\pi D_{ij} }{2} \\ 
0 & \textrm{otherwise}
\end{cases}
$$
Transforming the distances in this way  ensures that $M_{ij}$ only contains distances between atoms in the top and bottom layers.  Distances between atoms in the same layer or distances between atoms in one layer and the curve are excluded because the geodesic distance between them $|S_i - S_j|$ will be less than half the circumference of a sphere with diameter $D_{ij}$.  

We next introduce two vectors $\mathbf{r}^t$ and $\mathbf{r}^b$. The first of these vectors measures the distances between each of the $\frac{N_s}{2}$ 'safe' atoms in the top layer and the nearest atom in the bottom layer. The second vector does a similar measurement for the 'safe' atoms in the bottom layer. The components of these vectors are set equal to the smallest non-zero element in the row of $\mathbf{M}$ that corresponds to the 'safe' atom of interest.  From these vectors we compute the means $\langle r^t \rangle$ and $\langle r^b \rangle$ for the distribution of distances\footnote{To make the calculation of these mean values more robust we also calculate the standard deviation $\sigma$. We assume that any distance that is more than $2\sigma$ from the mean is an outlier, discard it and then recalculate the mean without including these outliers}.  $r$ in figure \ref{fig:dist_diag}, the distance between the top and bottom layers, is then set equal to the smaller of these two means.

In calculating the vectors $\mathbf{r}^t$ and $\mathbf{r}^b$ we also determine the atom from the opposite layer that is nearest to each of the 'safe' atoms.  These determinations are used when we calculate $d$.  To calculate this quantity we select the set of 'safe' atoms with the smaller average $r$ value and compute the following quantity:
$$
d_i = S_m - S_i - S_j
$$
for each of them.  In this expression, $S_i$ is the $S$ coordinate of the $i$th safe atom and $S_j$ is the $S$ coordinate of the closest atom from the opposite layer, which was determined from the calculation of $r$. $S_m$ is the total length of the flattened graphene sheet.  As figure \ref{fig:dist_diag} shows, $d_i$ is thus what remains once the geodesic distance between atom $i$ and the edge of the sheet and the geodesic distance between atom closest to atom $i$ from the opposite sheet is subtracted from the total length of the sheet.  

To arrive at a final scalar value for $d$, an average over all the individual $d_i$ values is computed.  We once again remove any outliers when computing this mean using the method that we described when we explained how $r$ is calculated.  

The solid black line in the top panel of figure \ref{fig:fitexample} shows how the value of $d$ changes over the course of a simulation.  The result shown here is for a $300\times 150$ \AA\ graphene sheet, which was started off in a configuration which had $r_1=2$ \AA\, $r_2=30$ \AA\ and $\theta=0$.  You can see that the $d$ undergoes a series of damped oscillations as would be expected given the snapshots from the trajectory that are shown in figure \ref{fig:singfold}.  \citet{Ye_2015} have observed similar behaviours in systems composed of flat graphene sheets that are stacked on top of each other and have fitted the oscillations using a damped harmonic oscillator model.  The blue line in figure in the upper panel of figure \ref{fig:fitexample} shows what we obtain when we attempt to fit the data using a similar damped harmonic oscillator model.  You can clearly see that there is close fit to the data from the early parts of the simulation but that there is a large discrepancy between the model and the data for large $t$.  The difficulties in fitting the data using this simple model suggest that the frequency of oscillations changes as the simulation progresses.  We thus chose to refit the data (red line) using the five parameter model below:
\begin{equation}
    x(t) = Ae^{-\gamma \omega t}\cos\left(\omega e^{\beta\omega t}t+\phi\right).
    \label{Eq:xfit}
\end{equation}
$A$ and $\phi$ are the initial amplitude and a phase factor and will not be discussed further in this work.  $\omega$ is the initial frequency of the oscillations and $\gamma$ is the damping parameter.  The $\beta$ parameter allows us to describe the changes in frequency with time that were observed (the blue curves are fits using the model above with $\beta=0$).  Figure \ref{fig:fitexample} illustrates that you get a much closer fit to the data when this $\beta$ parameter is allowed to vary, which confirms that the oscillation period changes during the simulation. 

If the folded graphene system is really undergoing damped oscillations that can be described using the model above, the layers should be moving relative to each other at a velocities given by the following expression:
\begin{equation}
\begin{aligned}
    v(t) = & -A\omega e^{-\gamma \omega t} 
    \left(\gamma \cos\left(\omega e^{\beta \omega t}t+\phi\right) 
    \right. \\
    &+ \left.e^{\beta\omega t}(\beta\omega t+1)\sin\left(\omega e^{\beta\omega t}t+\phi\right)\right) .
\end{aligned}
    \label{Eq:vfit}
\end{equation}
This expression was obtained by differentiating equation \ref{Eq:xfit}.  To test that this expression can indeed be used to describe the velocity of the oscillations we stored the velocities of all the 'safe' atoms.  The average velocity for the 'safe' atoms that are initially in the top layer was then computed along with the average velocity for the 'safe' atoms that were initially in the bottom layer.  The black line in the bottom panel of figure \ref{fig:fitexample} shows how the difference between these two averages changes during the simulation.  You can see that this quantity also oscillates, which is what one would expect from the model in equation \ref{Eq:vfit}. More interestingly, the blue line shows the result of fitting the time series of velocity differences from the simulation to equation \ref{Eq:vfit} with $\beta=0$, while the red line shows the result that is obtained when $\beta$ is a free parameter.  Once again you see that a closer fit is obtained when a parameter is included in the model that allows oscillation period to change as the simulation progresses.   

We observe that the fold moves relatively rapidly to a geometry, which has the two layers separated by their equilibrium distance and that oscillations start once this initial relaxation phase is completed.  We thus argue that the $\gamma$, $\omega$ and $\beta$ parameters of our damped harmonic oscillator model are properties of the graphene and not properties of the initial configuration.  To demonstrate this fact we run multiple simulations with different initial values for $r_1$ and $r_2$.  All values of $r_1$ and $r_2$ that undergo folding according to figure \ref{fig:foldcheck} were used.  The values that we quote for the parameters in the rest of this paper are found by averaging over these simulations.  We also quote the variance for the distribution of parameters and hence show that the parameters values we obtain do not depend strongly on $r_1$ or $r_2$.  As discussed in the next section, the parameters of these models change in ways that can be easily explained when the shape of the graphene sheet is varied.

\section{Effect of sheet dimensions}

\begin{figure}
    \centering
    \includegraphics[scale=0.5]{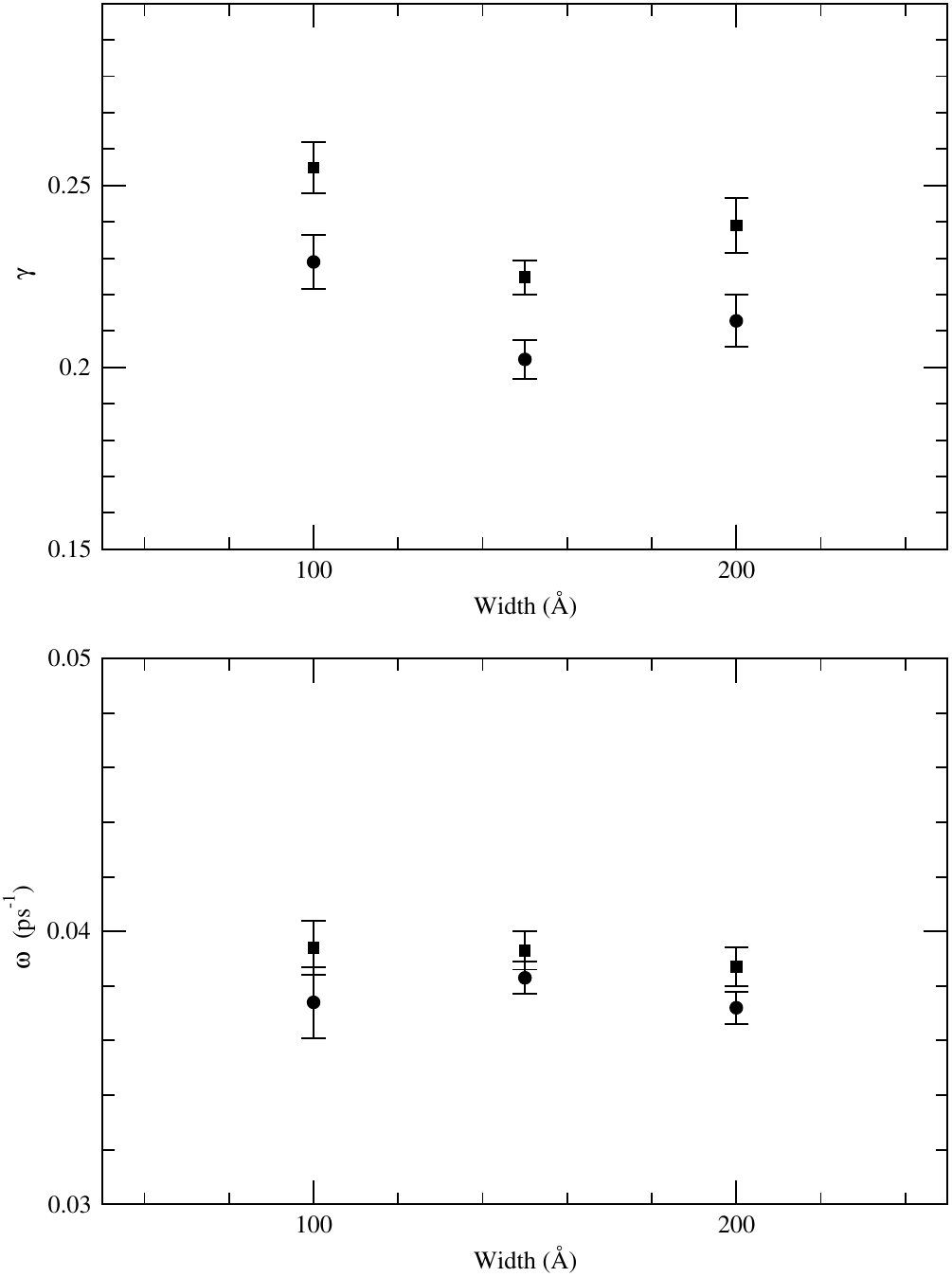}
    \caption{$\gamma$ (top figure) and $\omega$ (bottom figure) for sheets of various widths, a length of 300 \AA\ and $\theta=0^{\circ}$. Circles correspond to displacement data, squares from velocity data.}
    \label{fig:width_fits}
\end{figure}

The top and bottom panels of figure \ref{fig:width_fits} show the $\gamma$ and $\omega$ parameter values that were obtained for simulations of sheets with various widths that were all 300 \AA\ long.  Solid circles indicate the values that were obtained from fitting using equation \ref{Eq:xfit} and solid squares are the results that were obtained by fitting using equation \ref{Eq:vfit}.  You can see that neither $\omega$ or $\gamma$ depends strongly on the width of the sheet.  
This result is to be expected as we have not treated the edges of the graphene meaning that they are the only regions of high friction. In other words, the middle parts of the sheet only make a modest contribution to the total friction. Consequently, because increasing the width only increases the distance between the edges, $\gamma$ remains constant.
The distance the graphene sheet moves during each oscillation only depends on the length of the sheet. A wider sheet would also not be expected to move more rapidly than a narrower sheet. The fact that the lower panel of figure \ref{fig:width_fits} shows that the oscillation frequency, $\omega$, does not depend on the length of the sheet is thus not surprising.  

\begin{figure}
    \centering
    \includegraphics[scale=0.5]{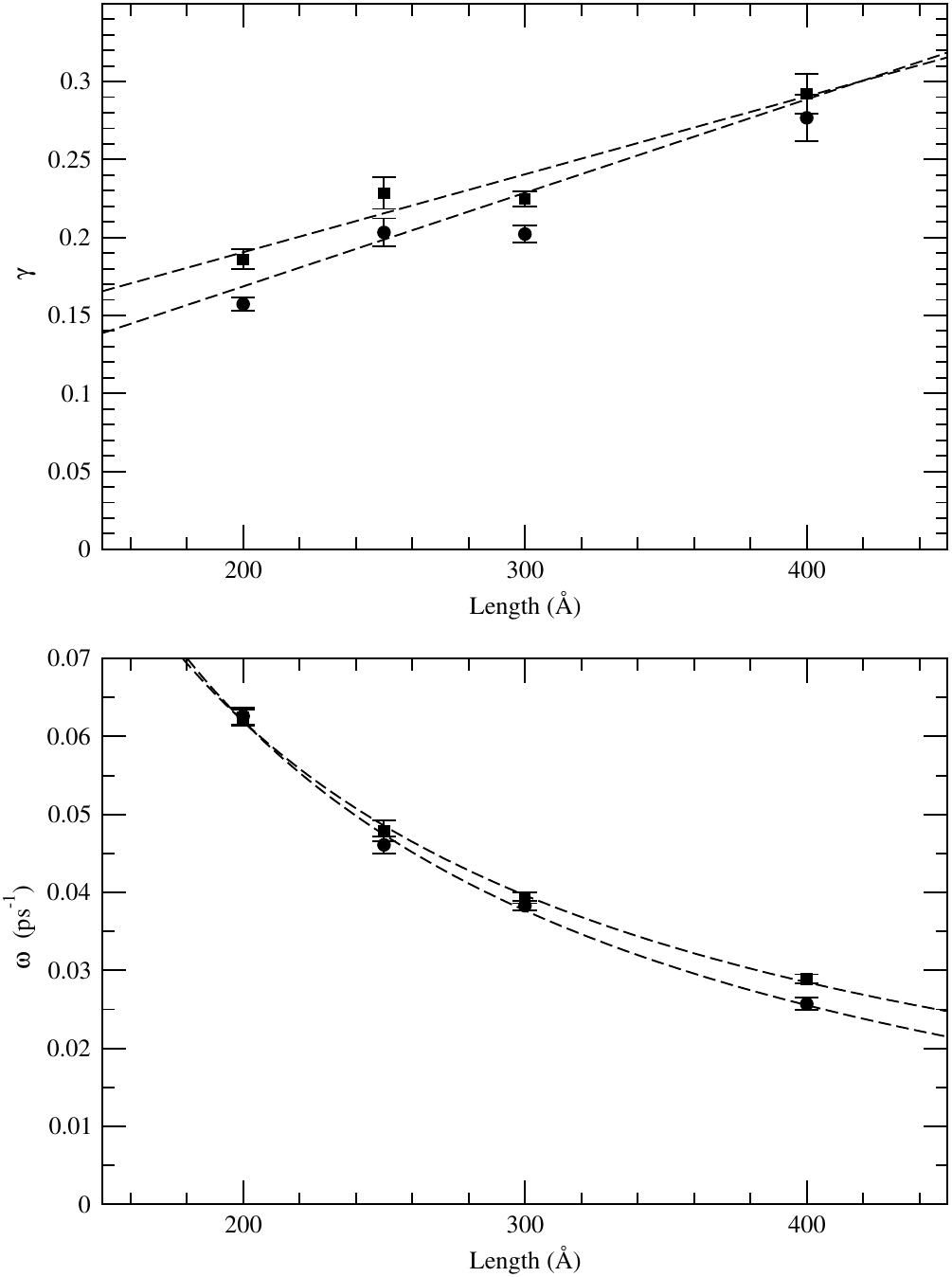}
    \caption{$\gamma$ (top figure) and $\omega$ (bottom figure) for sheets of various lengths, a width of 150 \AA\ and $\theta=0^{\circ}$. Legend the same as in Fig. \ref{fig:width_fits}. Dashed lines in the top figure corresponds to displacement and velocity data fitted to the formula $\gamma = M L +C$. Dashed lines in bottom figure are the same but for a fitting formula $\omega = \frac{M}{L}+C$.}
    \label{fig:length_fits}
\end{figure}

The top and bottom panels of figure \ref{fig:length_fits} show the $\gamma$ and $\omega$ parameter values that were obtained for simulations of sheets with various lengths that were all 150 \AA\ wide.  Once again, solid circles and solid squares are the results obtained using equations \ref{Eq:xfit} and \ref{Eq:vfit}.  The bottom panel of figure \ref{fig:length_fits} shows that $\omega$ decreases as the length of the sheet increases.  Furthermore, the dashed lines in this panel were generated by fitting these $\omega$ values to a function of the form $\frac{M}{L} + C$ where $L$ is the length of the sheet and $M$ and $C$ are parameters.  You can see that this function fits the data on $\omega$ reasonably closely. We believe that $\omega$ decreases with $\frac{1}{L}$ because when the sheet is longer it moves a greater distance during each oscillation.  As the velocity of the sheet does not increase when the sheet gets larger, completing a single oscillation takes longer when the sheet is larger.  

The top panel of figure \ref{fig:length_fits} shows that the $\gamma$ parameter increases as the sheet grows longer.  Furthermore, the increase in $\gamma$ is reasonably well described by a linear function of the length of the sheet.  This result is easy to explain.  Increasing the length of the sheet increases the lengths of the edges of the graphene. The majority of the friction between the two layers is known to come from interactions between the edges in the top and bottom layers of the sliding graphene sheet.  $\gamma$ should thus increase when the length of the sheet is increased and should remain the same (as we saw in the top panel of figure \ref{fig:width_fits}) when the width of the sheet is changed. 

\begin{figure}
    \centering
    \includegraphics[scale=0.35]{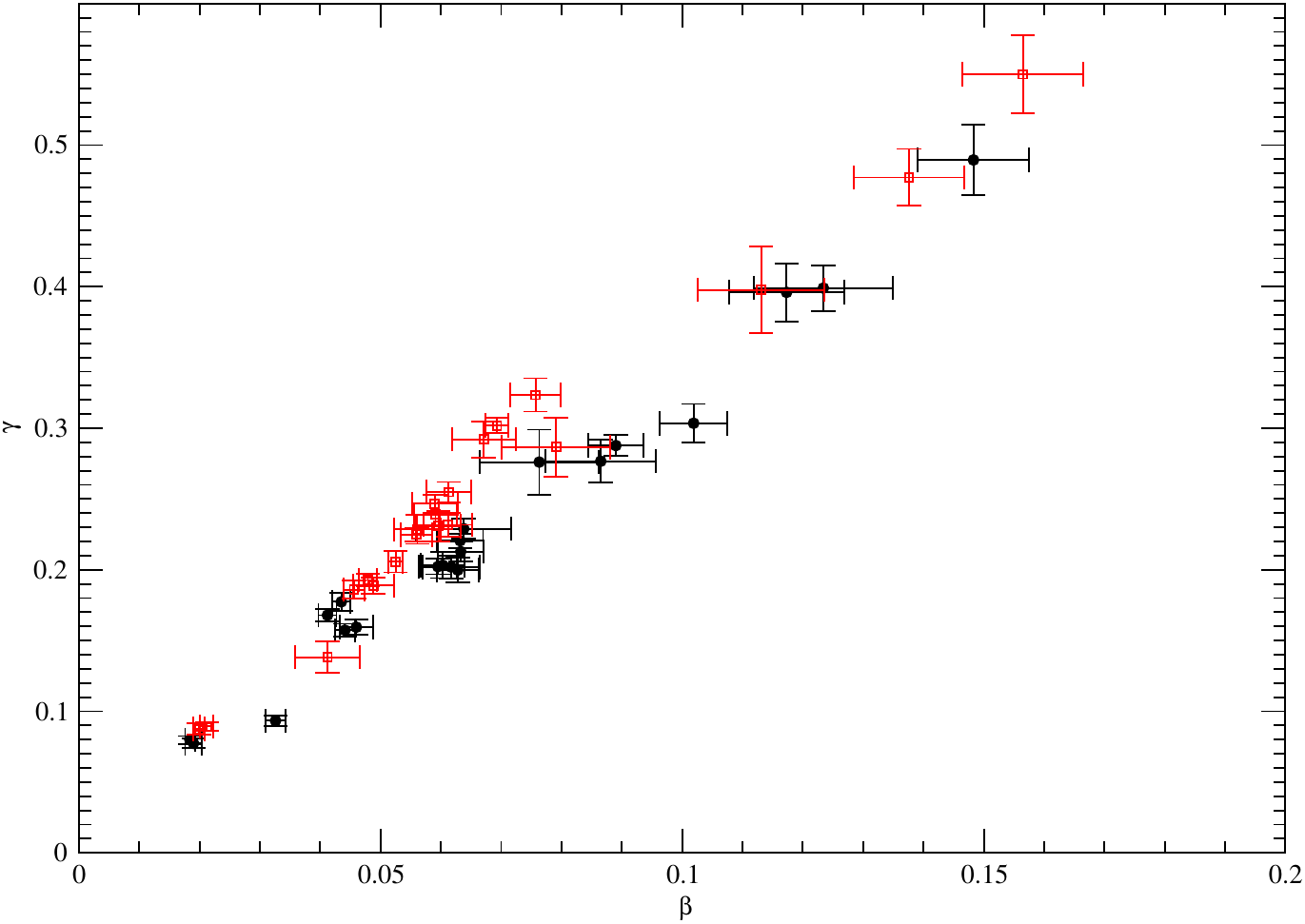}
    \caption{$\gamma$ vs $\beta$ for various fits of different sized sheets. Black circles corresponds to parameters from fitting of displacement data, red squares correspond to fitting with velocity data.}
    \label{fig:gammabeta}
\end{figure}

Figure \ref{fig:gammabeta} shows a plot of the $\gamma$ parameter against the $\beta$ parameter for all the simulations that were reported in figure \ref{fig:width_fits}, \ref{fig:length_fits} and \ref{fig:angle_gamma} (which is shown in a later section).  You can clearly see that there is a linear relationship between the $\gamma$ and $\beta$ parameters of the model.  We thus argue that these two parameters are  both providing equivalent information on the friction between the two layers.  In the remainder of the paper we thus report $\gamma$ values only as the same conclusion would be reached through an analysis of $\beta$.

\section{Effect of twist angle}

The previous section showed that the $\omega$ and $\beta$ parameters that emerge from our model are not particularly interesting.  There is a linear relationship between $\omega$ and the length of the sheet and the value of $\beta$ is proportional to the value of $\gamma$.  For the remainder of this paper we will thus only consider the $\gamma$ parameters that emerge from our model fitting.  

\begin{figure}
    \centering
    \includegraphics[scale=0.35]{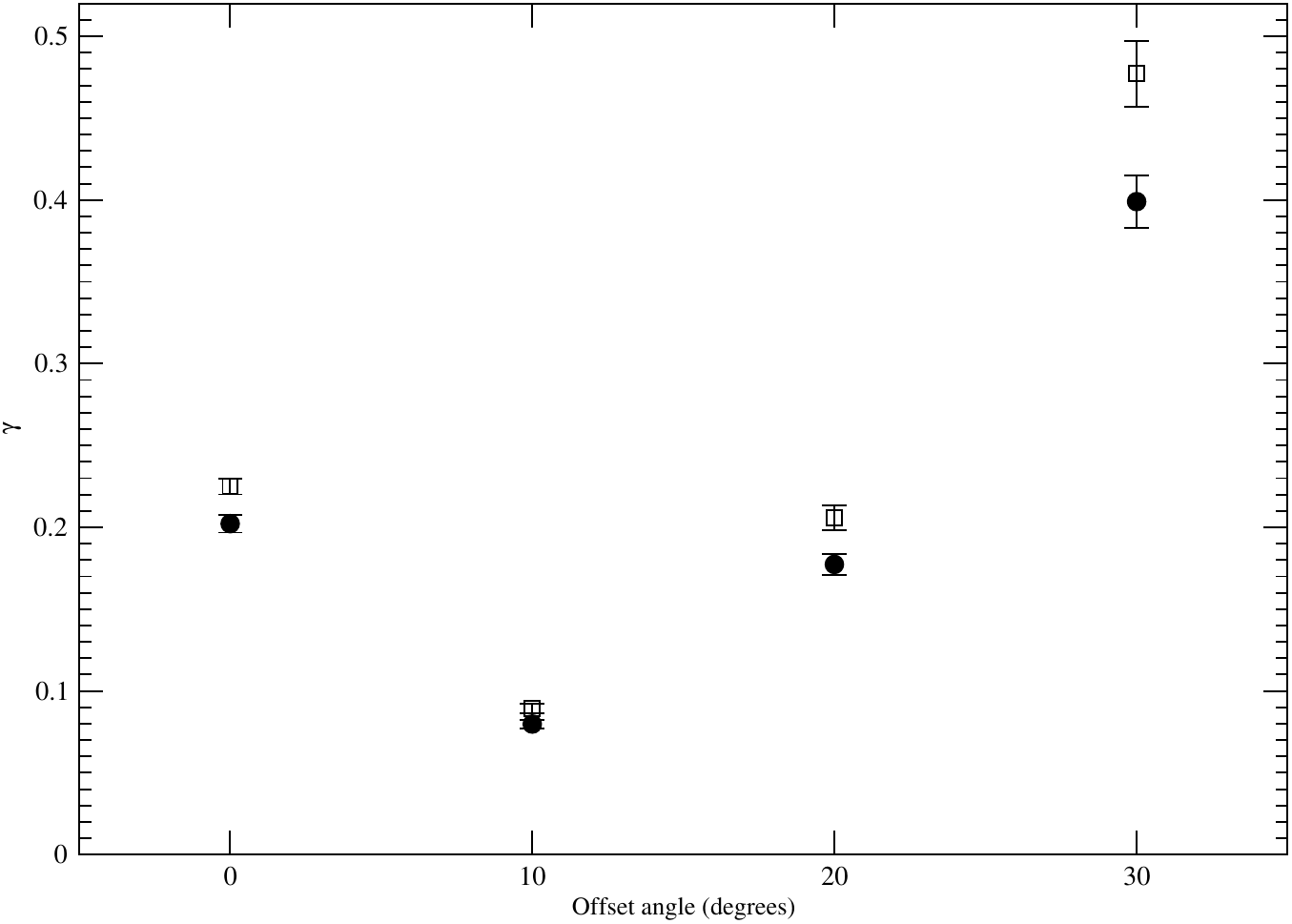}
    \caption{$\gamma$ vs $\theta$ for sheet dimensions 300 \AA\  by 150 \AA. Legend the same as for Fig. \ref{fig:width_fits}.}
    \label{fig:angle_gamma}
\end{figure}


Figure \ref{fig:angle_gamma} shows how the $\gamma$ parameter of the fit changes with the twist angle of the initial geometry. As shown in figure \ref{fig:layout}, this twist angle measures the angle between the $x$ axis of the lab frame and the armchair direction of the graphene sheet.  Figure \ref{fig:angle_gamma} shows that $\gamma$ is small when this angle is 10 or 20 degrees and that it is large when the angle is 0 or 30 degrees. This result makes physical sense as the two graphene layers are commensurately stacked when the angle is 0 or 30 degrees.  The sort of non-commensurate stacking layers that is known to give rise to the super lubricious sliding \cite{Dienwiebel_lubricity_04} is only present for the other two angles.

When the angle between the two graphene layers is 0 degrees the sliding (long) edge of the graphene fold has a arm-chair termination.  By contrast when the angle is 30 degrees this edge has an zig-zag termination. Our results thus suggest that arm-chair edges slide more easily against each other than zig-zag.

The results shown in figure \ref{fig:angle_gamma} were obtained from simulations of a $300 \times 150$ \AA\ graphene sheet. When we ran simulations with other sizes of graphene sheet the relative magnitude of the $\gamma$ parameters for different twist angles stayed the same.

\section{Effect of thermostat}

\begin{table}
    \centering
    \begin{tabular}{c|c|c}
       & $\gamma_x$ & $\gamma_v$ 
       \\ \hline
       $\theta=0$ \\
       NVT & 0.2022 $\pm$ 0.0054 & 0.2248 $\pm$ 0.0047 
       \\
       NVE & 0.1677 $\pm$ 0.0081 & 0.1939 $\pm$ 0.0056 
       \\
        \hline
        $\theta=10$\\
        NVT & 0.0798 $\pm$ 0.0027 & 0.0893 $\pm$ 0.0029 
        \\
        NVE & 0.0816 $\pm$ 0.0045 & 0.0929 $\pm$ 0.0061 
        \\ \hline
       $\theta=20$ \\
       NVT & 0.1773 $\pm$ 0.0062 & 0.2057 $\pm$ 0.0075 
       \\
       NVE & 0.1883 $\pm$ 0.0108 & 0.2187 $\pm$ 0.0128 
       \\ \hline
       $\theta=30$ \\
       NVT & 0.3990 $\pm$ 0.0160 & 0.4772 $\pm$ 0.0201 
       \\
       NVE & 0.3516 $\pm$ 0.0227 & 0.4209 $\pm$ 0.0269 
       \\ \hline
    \end{tabular}
    \caption{Fitting parameter $\gamma$ from the NVT and NVE ensembles}
    \label{tab:param_exam}
\end{table}
The simulations described in the previous sections were run in the NVT ensemble so a thermostat acts upon the atomic velocities.  To check that this thermostat is not dissipating energy from the oscillations we performed NVE simulations of the folded sheet.  A short equilibration in the NVT ensemble of 5 ps was required to move the system away from the artificial, initial configuration, which was very high in energy.  Results for the $\gamma$ parameters that were obtained from fitting the position and velocity curves using equations \ref{Eq:xfit} and \ref{Eq:vfit} for simulations run in both the NVT and NVE ensemble are provided in table \ref{tab:param_exam}. For most of the geometries considered the $\gamma$ parameters that emerge from the NVT simulations are within the numerical errors of the estimates that emerge from the NVE simulations. Even when larger discrepancies were identified we found that any difference in the values $\gamma \omega$ - a quantity that is not unit less and that has units of inverse time - for the NVE and NVT simulations were within statistical errors. We thus ascribe any apparently-statistically-significant differences in the $\gamma$ values for NVT and NVE simulations in table \ref{tab:param_exam} to issues with the fitting procedure and concluded that the thermostat is not dissipating energy from the oscillations.       


\section{Effect of temperature}

\begin{figure}
    \includegraphics[scale=0.35]{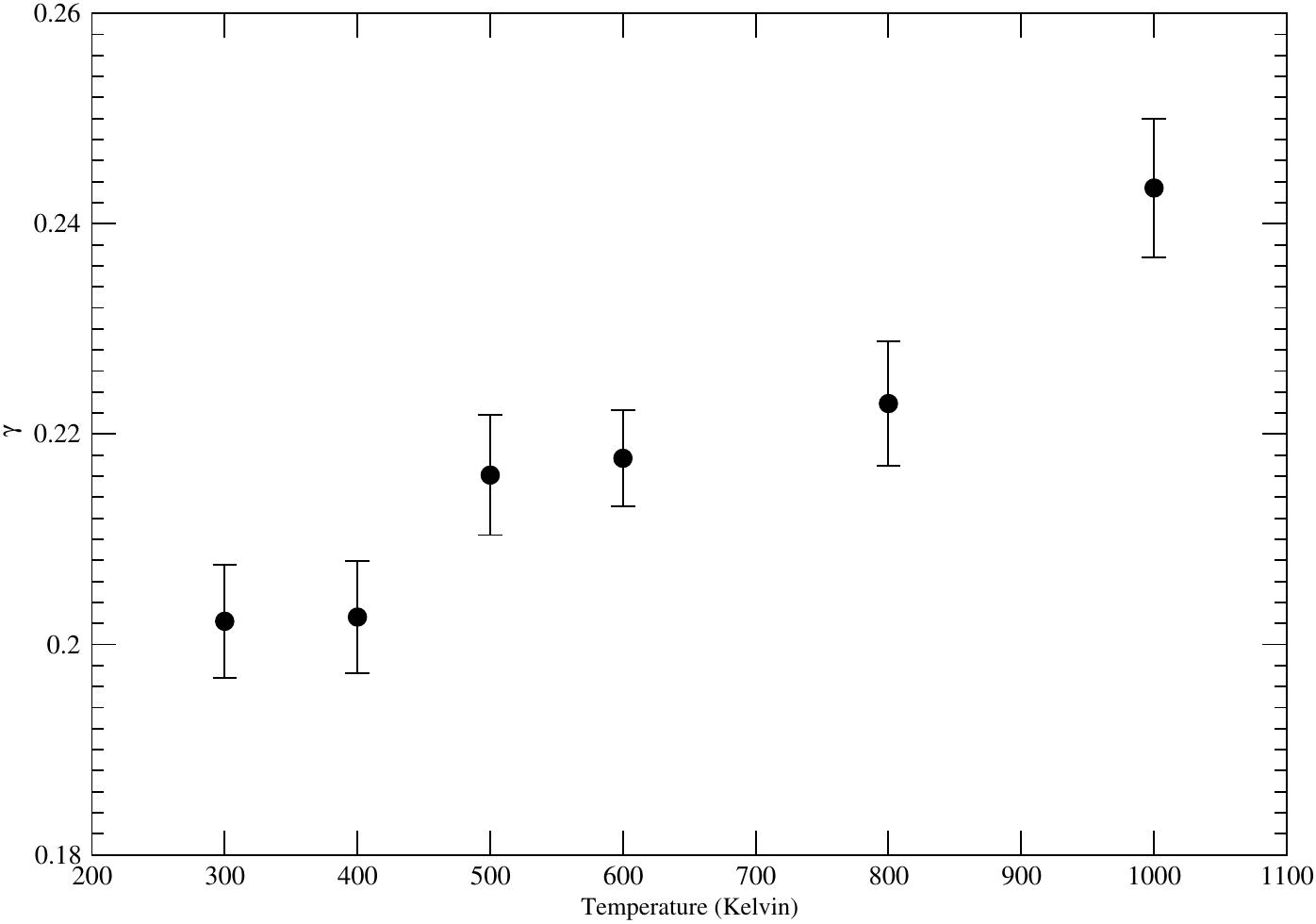}
    \caption{\label{fig:temp} $\gamma$ vs temperature for sheet dimensions 300 \AA\ by 150 \AA\ sheet and $\theta=0^{\circ}$.}
\end{figure}

Figure \ref{fig:temp} shows how the fitted $\gamma$ values from the model change with temperature.  This figure was generated by performing simulations on a 300$\times$150 \AA\ graphene sheet with a twist angle of 0 degrees. 

The level of variation in the damping parameter in figure \ref{fig:temp} is much lower than the variation that you see in figure \ref{fig:angle_gamma}. In other words, the relative orientation of the two layers has a much greater effect on the friction than the temperature.  Figure \ref{fig:temp} shows, however, that the friction appears to increase with temperature. \citet{Yang_13_lubricity} suggest that friction will increase with temperature as the graphene sheet fluctuates more at higher temperature.  These increased fluctuations make the surface appear rougher and thus more strongly suppress sliding motions.  Our simulations would suggest that this is a very small effect, however.  Furthermore, when we plot the parameters that emerge from fitting using equation \ref{Eq:vfit} we do not see the same monotonic increase in $\gamma$ with temperature.       



\section{Conclusions}

We have successfully developed a procedure to generate a folded sheet of graphene of arbitrary size and fold orientation, have used MD simulations to observe the sliding behaviour as the sheet settles to a fully-folded state and have developed an oscillator model to describe this sliding behaviour.
This model was then used to characterise the friction between the top and bottom layers of the folded sheet in terms of the decay parameter $\gamma$.
While this decay parameter is not an absolute measure of the friction of sliding graphene, examining the changes in $\gamma$ in different situations provides better qualitative understanding of how friction behaves at this level.
When varying the dimensions of the sheet, it was found that $\gamma$ scaled near linearly with sheet length, but had virtually no dependence of width.
This implies the greatest source of friction comes from the edges of the sheet, which is a very reasonable as nothing has been done to treat the edge atoms in our simulations.
Altering the offset angle $\theta$ shows that there is a significant drop in friction when going from commensurate stacking to non-commensurate stacking.
Finally, increasing the temperature causes a modest increase in sliding friction, which may be due to the 'roughness' of the sheet increasing with increased thermal fluctuation.

Our results show that the procedure developed here to model and analyse the sliding friction of a folded sheet of graphene produces results which are consistent with our current understanding of friction in these systems.
These results should provide some insights into the formation and propagation of kirigami ribbons in graphene observed by \citet{Cross_16}.
In order to better compare with experiment, the procedure will need to be updated in order to include a substrate, tearing mechanisms or even impurities in the system. Fortunately, the procedure is set up in such a way that these updates should not be a significant undertaking.
Our next priority is to include a substrate in the simulation in order to better delineate the affects of graphene-graphene and graphene-substrate interactions.

\begin{acknowledgments}
This work was supported by a research grant from the Department for the Economy Northern Ireland under the US-Ireland R\&D Partnership Programme. The computing resources used for this publication were from the Northern Ireland High Performance Computing (NI-HPC) service funded by EPSRC (EP/T022175). We thank Robert Carpick, Graham Cross, Ashlie Martini, David Wilkins and Myrta Gr{\"u}ning for useful discussions.
\end{acknowledgments}

\section*{Data Availability Statement}
The data that supports the findings of this study are available from the corresponding author upon reasonable request. 
Scripts for generating folded graphene sheets, input files for running on LAMMPS and scripts used to extract data from the resulting simulations as described in section \ref{Sec:fric} will be available on \href{https://github.com/crawlins5/Graphene_folding.git}{Github}.

\nocite{*}
\bibliography{references}

\begin{thebibliography}{30}%
\makeatletter
\providecommand \@ifxundefined [1]{%
 \@ifx{#1\undefined}
}%
\providecommand \@ifnum [1]{%
 \ifnum #1\expandafter \@firstoftwo
 \else \expandafter \@secondoftwo
 \fi
}%
\providecommand \@ifx [1]{%
 \ifx #1\expandafter \@firstoftwo
 \else \expandafter \@secondoftwo
 \fi
}%
\providecommand \natexlab [1]{#1}%
\providecommand \enquote  [1]{``#1''}%
\providecommand \bibnamefont  [1]{#1}%
\providecommand \bibfnamefont [1]{#1}%
\providecommand \citenamefont [1]{#1}%
\providecommand \href@noop [0]{\@secondoftwo}%
\providecommand \href [0]{\begingroup \@sanitize@url \@href}%
\providecommand \@href[1]{\@@startlink{#1}\@@href}%
\providecommand \@@href[1]{\endgroup#1\@@endlink}%
\providecommand \@sanitize@url [0]{\catcode `\\12\catcode `\$12\catcode
  `\&12\catcode `\#12\catcode `\^12\catcode `\_12\catcode `\%12\relax}%
\providecommand \@@startlink[1]{}%
\providecommand \@@endlink[0]{}%
\providecommand \url  [0]{\begingroup\@sanitize@url \@url }%
\providecommand \@url [1]{\endgroup\@href {#1}{\urlprefix }}%
\providecommand \urlprefix  [0]{URL }%
\providecommand \Eprint [0]{\href }%
\providecommand \doibase [0]{http://dx.doi.org/}%
\providecommand \selectlanguage [0]{\@gobble}%
\providecommand \bibinfo  [0]{\@secondoftwo}%
\providecommand \bibfield  [0]{\@secondoftwo}%
\providecommand \translation [1]{[#1]}%
\providecommand \BibitemOpen [0]{}%
\providecommand \bibitemStop [0]{}%
\providecommand \bibitemNoStop [0]{.\EOS\space}%
\providecommand \EOS [0]{\spacefactor3000\relax}%
\providecommand \BibitemShut  [1]{\csname bibitem#1\endcsname}%
\let\auto@bib@innerbib\@empty
\bibitem [{\citenamefont {Zhu}\ \emph {et~al.}(2010)\citenamefont {Zhu},
  \citenamefont {Murali}, \citenamefont {Cai}, \citenamefont {Li},
  \citenamefont {Suk}, \citenamefont {Potts},\ and\ \citenamefont
  {Ruoff}}]{Yanwu_10}%
  \BibitemOpen
  \bibfield  {author} {\bibinfo {author} {\bibfnamefont {Y.}~\bibnamefont
  {Zhu}}, \bibinfo {author} {\bibfnamefont {S.}~\bibnamefont {Murali}},
  \bibinfo {author} {\bibfnamefont {W.}~\bibnamefont {Cai}}, \bibinfo {author}
  {\bibfnamefont {X.}~\bibnamefont {Li}}, \bibinfo {author} {\bibfnamefont
  {J.~W.}\ \bibnamefont {Suk}}, \bibinfo {author} {\bibfnamefont {J.~R.}\
  \bibnamefont {Potts}}, \ and\ \bibinfo {author} {\bibfnamefont {R.~S.}\
  \bibnamefont {Ruoff}},\ }\bibfield  {title} {\enquote {\bibinfo {title}
  {Graphene and graphene oxide: Synthesis, properties, and applications},}\
  }\href {\doibase https://doi.org/10.1002/adma.201001068} {\bibfield
  {journal} {\bibinfo  {journal} {Advanced Materials}\ }\textbf {\bibinfo
  {volume} {22}},\ \bibinfo {pages} {3906--3924} (\bibinfo {year} {2010})},\
  \Eprint
  {http://arxiv.org/abs/https://onlinelibrary.wiley.com/doi/pdf/10.1002/adma.201001068}
  {https://onlinelibrary.wiley.com/doi/pdf/10.1002/adma.201001068} \BibitemShut
  {NoStop}%
\bibitem [{\citenamefont {Castro~Neto}\ \emph {et~al.}(2009)\citenamefont
  {Castro~Neto}, \citenamefont {Guinea}, \citenamefont {Peres}, \citenamefont
  {Novoselov},\ and\ \citenamefont {Geim}}]{Neto_09}%
  \BibitemOpen
  \bibfield  {author} {\bibinfo {author} {\bibfnamefont {A.~H.}\ \bibnamefont
  {Castro~Neto}}, \bibinfo {author} {\bibfnamefont {F.}~\bibnamefont {Guinea}},
  \bibinfo {author} {\bibfnamefont {N.~M.~R.}\ \bibnamefont {Peres}}, \bibinfo
  {author} {\bibfnamefont {K.~S.}\ \bibnamefont {Novoselov}}, \ and\ \bibinfo
  {author} {\bibfnamefont {A.~K.}\ \bibnamefont {Geim}},\ }\bibfield  {title}
  {\enquote {\bibinfo {title} {The electronic properties of graphene},}\ }\href
  {\doibase 10.1103/RevModPhys.81.109} {\bibfield  {journal} {\bibinfo
  {journal} {Rev. Mod. Phys.}\ }\textbf {\bibinfo {volume} {81}},\ \bibinfo
  {pages} {109--162} (\bibinfo {year} {2009})}\BibitemShut {NoStop}%
\bibitem [{\citenamefont {Viculis}, \citenamefont {Mack},\ and\ \citenamefont
  {Kaner}(2003)}]{Viculis_03}%
  \BibitemOpen
  \bibfield  {author} {\bibinfo {author} {\bibfnamefont {L.~M.}\ \bibnamefont
  {Viculis}}, \bibinfo {author} {\bibfnamefont {J.~J.}\ \bibnamefont {Mack}}, \
  and\ \bibinfo {author} {\bibfnamefont {R.~B.}\ \bibnamefont {Kaner}},\
  }\bibfield  {title} {\enquote {\bibinfo {title} {A chemical route to carbon
  nanoscrolls},}\ }\href {\doibase 10.1126/science.1078842} {\bibfield
  {journal} {\bibinfo  {journal} {Science}\ }\textbf {\bibinfo {volume}
  {299}},\ \bibinfo {pages} {1361} (\bibinfo {year} {2003})}\BibitemShut
  {NoStop}%
\bibitem [{\citenamefont {Pereira~Junior}\ and\ \citenamefont
  {Ribeiro~Junior}(2021)}]{Pereira_21}%
  \BibitemOpen
  \bibfield  {author} {\bibinfo {author} {\bibfnamefont {M.~L.}\ \bibnamefont
  {Pereira~Junior}}\ and\ \bibinfo {author} {\bibfnamefont {L.~A.}\
  \bibnamefont {Ribeiro~Junior}},\ }\bibfield  {title} {\enquote {\bibinfo
  {title} {Self-folding and self-scrolling mechanisms of edge-deformed graphene
  sheets: a molecular dynamics study},}\ }\href {\doibase 10.1039/D1CP02117F}
  {\bibfield  {journal} {\bibinfo  {journal} {Phys. Chem. Chem. Phys.}\
  }\textbf {\bibinfo {volume} {23}},\ \bibinfo {pages} {15313--15318} (\bibinfo
  {year} {2021})}\BibitemShut {NoStop}%
\bibitem [{\citenamefont {Zhang}\ \emph {et~al.}(2010)\citenamefont {Zhang},
  \citenamefont {Xiao}, \citenamefont {Meng}, \citenamefont {Monroe},
  \citenamefont {Huang},\ and\ \citenamefont {Zuo}}]{Zhang_10}%
  \BibitemOpen
  \bibfield  {author} {\bibinfo {author} {\bibfnamefont {J.}~\bibnamefont
  {Zhang}}, \bibinfo {author} {\bibfnamefont {J.}~\bibnamefont {Xiao}},
  \bibinfo {author} {\bibfnamefont {X.}~\bibnamefont {Meng}}, \bibinfo {author}
  {\bibfnamefont {C.}~\bibnamefont {Monroe}}, \bibinfo {author} {\bibfnamefont
  {Y.}~\bibnamefont {Huang}}, \ and\ \bibinfo {author} {\bibfnamefont {J.-M.}\
  \bibnamefont {Zuo}},\ }\bibfield  {title} {\enquote {\bibinfo {title} {Free
  folding of suspended graphene sheets by random mechanical stimulation},}\
  }\href {\doibase 10.1103/PhysRevLett.104.166805} {\bibfield  {journal}
  {\bibinfo  {journal} {Phys. Rev. Lett.}\ }\textbf {\bibinfo {volume} {104}},\
  \bibinfo {pages} {166805} (\bibinfo {year} {2010})}\BibitemShut {NoStop}%
\bibitem [{\citenamefont {Dienwiebel}\ \emph {et~al.}(2004)\citenamefont
  {Dienwiebel}, \citenamefont {Verhoeven}, \citenamefont {Pradeep},
  \citenamefont {Frenken}, \citenamefont {Heimberg},\ and\ \citenamefont
  {Zandbergen}}]{Dienwiebel_lubricity_04}%
  \BibitemOpen
  \bibfield  {author} {\bibinfo {author} {\bibfnamefont {M.}~\bibnamefont
  {Dienwiebel}}, \bibinfo {author} {\bibfnamefont {G.~S.}\ \bibnamefont
  {Verhoeven}}, \bibinfo {author} {\bibfnamefont {N.}~\bibnamefont {Pradeep}},
  \bibinfo {author} {\bibfnamefont {J.~W.~M.}\ \bibnamefont {Frenken}},
  \bibinfo {author} {\bibfnamefont {J.~A.}\ \bibnamefont {Heimberg}}, \ and\
  \bibinfo {author} {\bibfnamefont {H.~W.}\ \bibnamefont {Zandbergen}},\
  }\bibfield  {title} {\enquote {\bibinfo {title} {Superlubricity of
  graphite},}\ }\href {\doibase 10.1103/PhysRevLett.92.126101} {\bibfield
  {journal} {\bibinfo  {journal} {Phys. Rev. Lett.}\ }\textbf {\bibinfo
  {volume} {92}},\ \bibinfo {pages} {126101} (\bibinfo {year}
  {2004})}\BibitemShut {NoStop}%
\bibitem [{\citenamefont {Liu}, \citenamefont {Grey},\ and\ \citenamefont
  {Zheng}(2014)}]{Yilun_14}%
  \BibitemOpen
  \bibfield  {author} {\bibinfo {author} {\bibfnamefont {Y.}~\bibnamefont
  {Liu}}, \bibinfo {author} {\bibfnamefont {F.}~\bibnamefont {Grey}}, \ and\
  \bibinfo {author} {\bibfnamefont {Q.}~\bibnamefont {Zheng}},\ }\bibfield
  {title} {\enquote {\bibinfo {title} {The high-speed sliding friction of
  graphene and novel routes to persistent superlubricity},}\ }\href {\doibase
  10.1038/srep04875} {\bibfield  {journal} {\bibinfo  {journal} {Scientific
  Reports}\ }\textbf {\bibinfo {volume} {4}},\ \bibinfo {pages} {4875}
  (\bibinfo {year} {2014})}\BibitemShut {NoStop}%
\bibitem [{\citenamefont {Annett}\ and\ \citenamefont
  {Cross}(2016)}]{Cross_16}%
  \BibitemOpen
  \bibfield  {author} {\bibinfo {author} {\bibfnamefont {J.}~\bibnamefont
  {Annett}}\ and\ \bibinfo {author} {\bibfnamefont {G.~L.~W.}\ \bibnamefont
  {Cross}},\ }\bibfield  {title} {\enquote {\bibinfo {title} {Self-assembly of
  graphene ribbons by spontaneous self-tearing and peeling from a substrate},}\
  }\href {\doibase 10.1038/nature18304} {\bibfield  {journal} {\bibinfo
  {journal} {Nature}\ }\textbf {\bibinfo {volume} {535}},\ \bibinfo {pages}
  {271--275} (\bibinfo {year} {2016})}\BibitemShut {NoStop}%
\bibitem [{\citenamefont {Fonseca}\ and\ \citenamefont
  {Galv{\~a}o}(2018{\natexlab{a}})}]{Fonseca2018SelftearingAS}%
  \BibitemOpen
  \bibfield  {author} {\bibinfo {author} {\bibfnamefont {A.~F.}\ \bibnamefont
  {Fonseca}}\ and\ \bibinfo {author} {\bibfnamefont {D.~S.}\ \bibnamefont
  {Galv{\~a}o}},\ }\bibfield  {title} {\enquote {\bibinfo {title} {Self-tearing
  and self-peeling of folded graphene nanoribbons},}\ }\href
  {https://api.semanticscholar.org/CorpusID:119398762} {\bibfield  {journal}
  {\bibinfo  {journal} {Carbon}\ } (\bibinfo {year}
  {2018}{\natexlab{a}})}\BibitemShut {NoStop}%
\bibitem [{\citenamefont {Fonseca}\ and\ \citenamefont
  {Galv{\~a}o}(2018{\natexlab{b}})}]{Fonseca_18}%
  \BibitemOpen
  \bibfield  {author} {\bibinfo {author} {\bibfnamefont {A.~F.}\ \bibnamefont
  {Fonseca}}\ and\ \bibinfo {author} {\bibfnamefont {D.~S.}\ \bibnamefont
  {Galv{\~a}o}},\ }\bibfield  {title} {\enquote {\bibinfo {title} {Self-driven
  graphene tearing and peeling: A fully atomistic molecular dynamics
  investigation},}\ }\href {\doibase 10.1557/adv.2018.120} {\bibfield
  {journal} {\bibinfo  {journal} {MRS Advances}\ }\textbf {\bibinfo {volume}
  {3}},\ \bibinfo {pages} {463--468} (\bibinfo {year}
  {2018}{\natexlab{b}})}\BibitemShut {NoStop}%
\bibitem [{\citenamefont {He}, \citenamefont {Zhu},\ and\ \citenamefont
  {Wu}(2018)}]{He_18}%
  \BibitemOpen
  \bibfield  {author} {\bibinfo {author} {\bibfnamefont {Z.-Z.}\ \bibnamefont
  {He}}, \bibinfo {author} {\bibfnamefont {Y.-B.}\ \bibnamefont {Zhu}}, \ and\
  \bibinfo {author} {\bibfnamefont {H.-A.}\ \bibnamefont {Wu}},\ }\bibfield
  {title} {\enquote {\bibinfo {title} {Self-folding mechanics of graphene
  tearing and peeling from a substrate},}\ }\href {\doibase
  10.1007/s11467-018-0755-5} {\bibfield  {journal} {\bibinfo  {journal}
  {Frontiers of Physics}\ }\textbf {\bibinfo {volume} {13}},\ \bibinfo {pages}
  {138111} (\bibinfo {year} {2018})}\BibitemShut {NoStop}%
\bibitem [{\citenamefont {Yang}\ \emph {et~al.}(2013)\citenamefont {Yang},
  \citenamefont {Liu}, \citenamefont {Grey}, \citenamefont {Xu}, \citenamefont
  {Li}, \citenamefont {Liu}, \citenamefont {Urbakh}, \citenamefont {Cheng},\
  and\ \citenamefont {Zheng}}]{Yang_13_lubricity}%
  \BibitemOpen
  \bibfield  {author} {\bibinfo {author} {\bibfnamefont {J.}~\bibnamefont
  {Yang}}, \bibinfo {author} {\bibfnamefont {Z.}~\bibnamefont {Liu}}, \bibinfo
  {author} {\bibfnamefont {F.}~\bibnamefont {Grey}}, \bibinfo {author}
  {\bibfnamefont {Z.}~\bibnamefont {Xu}}, \bibinfo {author} {\bibfnamefont
  {X.}~\bibnamefont {Li}}, \bibinfo {author} {\bibfnamefont {Y.}~\bibnamefont
  {Liu}}, \bibinfo {author} {\bibfnamefont {M.}~\bibnamefont {Urbakh}},
  \bibinfo {author} {\bibfnamefont {Y.}~\bibnamefont {Cheng}}, \ and\ \bibinfo
  {author} {\bibfnamefont {Q.}~\bibnamefont {Zheng}},\ }\bibfield  {title}
  {\enquote {\bibinfo {title} {Observation of high-speed microscale
  superlubricity in graphite},}\ }\href {\doibase
  10.1103/PhysRevLett.110.255504} {\bibfield  {journal} {\bibinfo  {journal}
  {Phys. Rev. Lett.}\ }\textbf {\bibinfo {volume} {110}},\ \bibinfo {pages}
  {255504} (\bibinfo {year} {2013})}\BibitemShut {NoStop}%
\bibitem [{\citenamefont {Liu}\ \emph {et~al.}(2012{\natexlab{a}})\citenamefont
  {Liu}, \citenamefont {Yang}, \citenamefont {Grey}, \citenamefont {Liu},
  \citenamefont {Liu}, \citenamefont {Wang}, \citenamefont {Yang},
  \citenamefont {Cheng},\ and\ \citenamefont {Zheng}}]{Yang_12_lubricity}%
  \BibitemOpen
  \bibfield  {author} {\bibinfo {author} {\bibfnamefont {Z.}~\bibnamefont
  {Liu}}, \bibinfo {author} {\bibfnamefont {J.}~\bibnamefont {Yang}}, \bibinfo
  {author} {\bibfnamefont {F.}~\bibnamefont {Grey}}, \bibinfo {author}
  {\bibfnamefont {J.~Z.}\ \bibnamefont {Liu}}, \bibinfo {author} {\bibfnamefont
  {Y.}~\bibnamefont {Liu}}, \bibinfo {author} {\bibfnamefont {Y.}~\bibnamefont
  {Wang}}, \bibinfo {author} {\bibfnamefont {Y.}~\bibnamefont {Yang}}, \bibinfo
  {author} {\bibfnamefont {Y.}~\bibnamefont {Cheng}}, \ and\ \bibinfo {author}
  {\bibfnamefont {Q.}~\bibnamefont {Zheng}},\ }\bibfield  {title} {\enquote
  {\bibinfo {title} {Observation of microscale superlubricity in graphite},}\
  }\href {\doibase 10.1103/PhysRevLett.108.205503} {\bibfield  {journal}
  {\bibinfo  {journal} {Phys. Rev. Lett.}\ }\textbf {\bibinfo {volume} {108}},\
  \bibinfo {pages} {205503} (\bibinfo {year} {2012}{\natexlab{a}})}\BibitemShut
  {NoStop}%
\bibitem [{\citenamefont {Zheng}\ \emph {et~al.}(2008)\citenamefont {Zheng},
  \citenamefont {Jiang}, \citenamefont {Liu}, \citenamefont {Weng},
  \citenamefont {Lu}, \citenamefont {Xue}, \citenamefont {Zhu}, \citenamefont
  {Jiang}, \citenamefont {Wang},\ and\ \citenamefont
  {Peng}}]{Zheng_08_self_retract}%
  \BibitemOpen
  \bibfield  {author} {\bibinfo {author} {\bibfnamefont {Q.}~\bibnamefont
  {Zheng}}, \bibinfo {author} {\bibfnamefont {B.}~\bibnamefont {Jiang}},
  \bibinfo {author} {\bibfnamefont {S.}~\bibnamefont {Liu}}, \bibinfo {author}
  {\bibfnamefont {Y.}~\bibnamefont {Weng}}, \bibinfo {author} {\bibfnamefont
  {L.}~\bibnamefont {Lu}}, \bibinfo {author} {\bibfnamefont {Q.}~\bibnamefont
  {Xue}}, \bibinfo {author} {\bibfnamefont {J.}~\bibnamefont {Zhu}}, \bibinfo
  {author} {\bibfnamefont {Q.}~\bibnamefont {Jiang}}, \bibinfo {author}
  {\bibfnamefont {S.}~\bibnamefont {Wang}}, \ and\ \bibinfo {author}
  {\bibfnamefont {L.}~\bibnamefont {Peng}},\ }\bibfield  {title} {\enquote
  {\bibinfo {title} {Self-retracting motion of graphite microflakes},}\ }\href
  {\doibase 10.1103/PhysRevLett.100.067205} {\bibfield  {journal} {\bibinfo
  {journal} {Phys. Rev. Lett.}\ }\textbf {\bibinfo {volume} {100}},\ \bibinfo
  {pages} {067205} (\bibinfo {year} {2008})}\BibitemShut {NoStop}%
\bibitem [{\citenamefont {Ng}\ \emph {et~al.}(2012)\citenamefont {Ng},
  \citenamefont {Lau}, \citenamefont {Bernados-Chamagne}, \citenamefont {Liu},
  \citenamefont {Sheridan},\ and\ \citenamefont {Tan}}]{Ng_12_Self_retract}%
  \BibitemOpen
  \bibfield  {author} {\bibinfo {author} {\bibfnamefont {T.~W.}\ \bibnamefont
  {Ng}}, \bibinfo {author} {\bibfnamefont {C.~Y.}\ \bibnamefont {Lau}},
  \bibinfo {author} {\bibfnamefont {E.}~\bibnamefont {Bernados-Chamagne}},
  \bibinfo {author} {\bibfnamefont {J.~Z.}\ \bibnamefont {Liu}}, \bibinfo
  {author} {\bibfnamefont {J.}~\bibnamefont {Sheridan}}, \ and\ \bibinfo
  {author} {\bibfnamefont {N.}~\bibnamefont {Tan}},\ }\bibfield  {title}
  {\enquote {\bibinfo {title} {Graphite flake self-retraction response based on
  potential seeking},}\ }\href {\doibase 10.1186/1556-276X-7-185} {\bibfield
  {journal} {\bibinfo  {journal} {Nanoscale Research Letters}\ }\textbf
  {\bibinfo {volume} {7}},\ \bibinfo {pages} {185} (\bibinfo {year}
  {2012})}\BibitemShut {NoStop}%
\bibitem [{\citenamefont {Lebedeva}\ \emph {et~al.}(2011)\citenamefont
  {Lebedeva}, \citenamefont {Knizhnik}, \citenamefont {Popov}, \citenamefont
  {Lozovik},\ and\ \citenamefont {Potapkin}}]{Lebedeva_11}%
  \BibitemOpen
  \bibfield  {author} {\bibinfo {author} {\bibfnamefont {I.~V.}\ \bibnamefont
  {Lebedeva}}, \bibinfo {author} {\bibfnamefont {A.~A.}\ \bibnamefont
  {Knizhnik}}, \bibinfo {author} {\bibfnamefont {A.~M.}\ \bibnamefont {Popov}},
  \bibinfo {author} {\bibfnamefont {Y.~E.}\ \bibnamefont {Lozovik}}, \ and\
  \bibinfo {author} {\bibfnamefont {B.~V.}\ \bibnamefont {Potapkin}},\
  }\bibfield  {title} {\enquote {\bibinfo {title} {Interlayer interaction and
  relative vibrations of bilayer graphene},}\ }\href {\doibase
  10.1039/C0CP02614J} {\bibfield  {journal} {\bibinfo  {journal} {Phys. Chem.
  Chem. Phys.}\ }\textbf {\bibinfo {volume} {13}},\ \bibinfo {pages}
  {5687--5695} (\bibinfo {year} {2011})}\BibitemShut {NoStop}%
\bibitem [{\citenamefont {Xu}\ \emph {et~al.}(2013)\citenamefont {Xu},
  \citenamefont {Li}, \citenamefont {Yakobson},\ and\ \citenamefont
  {Ding}}]{Xu_13_commensurate}%
  \BibitemOpen
  \bibfield  {author} {\bibinfo {author} {\bibfnamefont {Z.}~\bibnamefont
  {Xu}}, \bibinfo {author} {\bibfnamefont {X.}~\bibnamefont {Li}}, \bibinfo
  {author} {\bibfnamefont {B.~I.}\ \bibnamefont {Yakobson}}, \ and\ \bibinfo
  {author} {\bibfnamefont {F.}~\bibnamefont {Ding}},\ }\bibfield  {title}
  {\enquote {\bibinfo {title} {Interaction between graphene layers and the
  mechanisms of graphite{'}s superlubricity and self-retraction},}\ }\href
  {\doibase 10.1039/C3NR01854G} {\bibfield  {journal} {\bibinfo  {journal}
  {Nanoscale}\ }\textbf {\bibinfo {volume} {5}},\ \bibinfo {pages} {6736--6741}
  (\bibinfo {year} {2013})}\BibitemShut {NoStop}%
\bibitem [{\citenamefont {Popov}\ \emph
  {et~al.}(2011{\natexlab{a}})\citenamefont {Popov}, \citenamefont {Lebedeva},
  \citenamefont {Knizhnik}, \citenamefont {Lozovik},\ and\ \citenamefont
  {Potapkin}}]{Popov_11_commensurate}%
  \BibitemOpen
  \bibfield  {author} {\bibinfo {author} {\bibfnamefont {A.~M.}\ \bibnamefont
  {Popov}}, \bibinfo {author} {\bibfnamefont {I.~V.}\ \bibnamefont {Lebedeva}},
  \bibinfo {author} {\bibfnamefont {A.~A.}\ \bibnamefont {Knizhnik}}, \bibinfo
  {author} {\bibfnamefont {Y.~E.}\ \bibnamefont {Lozovik}}, \ and\ \bibinfo
  {author} {\bibfnamefont {B.~V.}\ \bibnamefont {Potapkin}},\ }\bibfield
  {title} {\enquote {\bibinfo {title} {Molecular dynamics simulation of the
  self-retracting motion of a graphene flake},}\ }\href {\doibase
  10.1103/PhysRevB.84.245437} {\bibfield  {journal} {\bibinfo  {journal} {Phys.
  Rev. B}\ }\textbf {\bibinfo {volume} {84}},\ \bibinfo {pages} {245437}
  (\bibinfo {year} {2011}{\natexlab{a}})}\BibitemShut {NoStop}%
\bibitem [{\citenamefont {Popov}\ \emph
  {et~al.}(2011{\natexlab{b}})\citenamefont {Popov}, \citenamefont {Lebedeva},
  \citenamefont {Knizhnik}, \citenamefont {Lozovik},\ and\ \citenamefont
  {Potapkin}}]{Popov_11_commensurateb}%
  \BibitemOpen
  \bibfield  {author} {\bibinfo {author} {\bibfnamefont {A.~M.}\ \bibnamefont
  {Popov}}, \bibinfo {author} {\bibfnamefont {I.~V.}\ \bibnamefont {Lebedeva}},
  \bibinfo {author} {\bibfnamefont {A.~A.}\ \bibnamefont {Knizhnik}}, \bibinfo
  {author} {\bibfnamefont {Y.~E.}\ \bibnamefont {Lozovik}}, \ and\ \bibinfo
  {author} {\bibfnamefont {B.~V.}\ \bibnamefont {Potapkin}},\ }\bibfield
  {title} {\enquote {\bibinfo {title} {Commensurate-incommensurate phase
  transition in bilayer graphene},}\ }\href {\doibase
  10.1103/PhysRevB.84.045404} {\bibfield  {journal} {\bibinfo  {journal} {Phys.
  Rev. B}\ }\textbf {\bibinfo {volume} {84}},\ \bibinfo {pages} {045404}
  (\bibinfo {year} {2011}{\natexlab{b}})}\BibitemShut {NoStop}%
\bibitem [{\citenamefont {Ye}\ \emph {et~al.}(2014)\citenamefont {Ye},
  \citenamefont {de-la Roza}, \citenamefont {Johnson},\ and\ \citenamefont
  {Martini}}]{Ye_2014}%
  \BibitemOpen
  \bibfield  {author} {\bibinfo {author} {\bibfnamefont {Z.}~\bibnamefont
  {Ye}}, \bibinfo {author} {\bibfnamefont {A.~O.}\ \bibnamefont {de-la Roza}},
  \bibinfo {author} {\bibfnamefont {E.~R.}\ \bibnamefont {Johnson}}, \ and\
  \bibinfo {author} {\bibfnamefont {A.}~\bibnamefont {Martini}},\ }\bibfield
  {title} {\enquote {\bibinfo {title} {The role of roughness-induced damping in
  the oscillatory motion of bilayer graphene},}\ }\href {\doibase
  10.1088/0957-4484/25/42/425703} {\bibfield  {journal} {\bibinfo  {journal}
  {Nanotechnology}\ }\textbf {\bibinfo {volume} {25}},\ \bibinfo {pages}
  {425703} (\bibinfo {year} {2014})}\BibitemShut {NoStop}%
\bibitem [{\citenamefont {Ye}\ \emph {et~al.}(2015)\citenamefont {Ye},
  \citenamefont {de-la Roza}, \citenamefont {Johnson},\ and\ \citenamefont
  {Martini}}]{Ye_2015}%
  \BibitemOpen
  \bibfield  {author} {\bibinfo {author} {\bibfnamefont {Z.}~\bibnamefont
  {Ye}}, \bibinfo {author} {\bibfnamefont {A.~O.}\ \bibnamefont {de-la Roza}},
  \bibinfo {author} {\bibfnamefont {E.~R.}\ \bibnamefont {Johnson}}, \ and\
  \bibinfo {author} {\bibfnamefont {A.}~\bibnamefont {Martini}},\ }\bibfield
  {title} {\enquote {\bibinfo {title} {Oscillatory motion in layered materials:
  graphene, boron nitride, and molybdenum disulfide},}\ }\href {\doibase
  10.1088/0957-4484/26/16/165701} {\bibfield  {journal} {\bibinfo  {journal}
  {Nanotechnology}\ }\textbf {\bibinfo {volume} {26}},\ \bibinfo {pages}
  {165701} (\bibinfo {year} {2015})}\BibitemShut {NoStop}%
\bibitem [{\citenamefont {van Duin}\ \emph {et~al.}(2001)\citenamefont {van
  Duin}, \citenamefont {Dasgupta}, \citenamefont {Lorant},\ and\ \citenamefont
  {Goddard}}]{ReaxFF_1}%
  \BibitemOpen
  \bibfield  {author} {\bibinfo {author} {\bibfnamefont {A.~C.~T.}\
  \bibnamefont {van Duin}}, \bibinfo {author} {\bibfnamefont {S.}~\bibnamefont
  {Dasgupta}}, \bibinfo {author} {\bibfnamefont {F.}~\bibnamefont {Lorant}}, \
  and\ \bibinfo {author} {\bibfnamefont {W.~A.}\ \bibnamefont {Goddard}},\
  }\bibfield  {title} {\enquote {\bibinfo {title} {Reaxff: A reactive force
  field for hydrocarbons},}\ }\href {\doibase 10.1021/jp004368u} {\bibfield
  {journal} {\bibinfo  {journal} {The Journal of Physical Chemistry A}\
  }\textbf {\bibinfo {volume} {105}},\ \bibinfo {pages} {9396--9409} (\bibinfo
  {year} {2001})}\BibitemShut {NoStop}%
\bibitem [{\citenamefont {Mueller}, \citenamefont {van Duin},\ and\
  \citenamefont {Goddard}(2010)}]{ReaxFF_2}%
  \BibitemOpen
  \bibfield  {author} {\bibinfo {author} {\bibfnamefont {J.~E.}\ \bibnamefont
  {Mueller}}, \bibinfo {author} {\bibfnamefont {A.~C.~T.}\ \bibnamefont {van
  Duin}}, \ and\ \bibinfo {author} {\bibfnamefont {W.~A.~I.}\ \bibnamefont
  {Goddard}},\ }\bibfield  {title} {\enquote {\bibinfo {title} {Development and
  validation of reaxff reactive force field for hydrocarbon chemistry catalyzed
  by nickel},}\ }\href {\doibase 10.1021/jp9035056} {\bibfield  {journal}
  {\bibinfo  {journal} {The Journal of Physical Chemistry C}\ }\textbf
  {\bibinfo {volume} {114}},\ \bibinfo {pages} {4939--4949} (\bibinfo {year}
  {2010})}\BibitemShut {NoStop}%
\bibitem [{\citenamefont {Aktulga}, \citenamefont {J.~C.~Fogarty},\ and\
  \citenamefont {Grama}(2012)}]{Aktulga12}%
  \BibitemOpen
  \bibfield  {author} {\bibinfo {author} {\bibfnamefont {H.~M.}\ \bibnamefont
  {Aktulga}}, \bibinfo {author} {\bibfnamefont {S.~A.~P.}\ \bibnamefont
  {J.~C.~Fogarty}}, \ and\ \bibinfo {author} {\bibfnamefont {A.~Y.}\
  \bibnamefont {Grama}},\ }\bibfield  {title} {\enquote {\bibinfo {title}
  {Parallel reactive molecular dynamics: Numerical methods and algorithmic
  techniques},}\ }\href@noop {} {\bibfield  {journal} {\bibinfo  {journal}
  {Parallel Computing}\ }\textbf {\bibinfo {volume} {38}},\ \bibinfo {pages}
  {245--259} (\bibinfo {year} {2012})}\BibitemShut {NoStop}%
\bibitem [{\citenamefont {Thompson}\ \emph {et~al.}(2022)\citenamefont
  {Thompson}, \citenamefont {Aktulga}, \citenamefont {Berger}, \citenamefont
  {Bolintineanu}, \citenamefont {Brown}, \citenamefont {Crozier}, \citenamefont
  {in~'t Veld}, \citenamefont {Kohlmeyer}, \citenamefont {Moore}, \citenamefont
  {Nguyen}, \citenamefont {Shan}, \citenamefont {Stevens}, \citenamefont
  {Tranchida}, \citenamefont {Trott},\ and\ \citenamefont {Plimpton}}]{LAMMPS}%
  \BibitemOpen
  \bibfield  {author} {\bibinfo {author} {\bibfnamefont {A.~P.}\ \bibnamefont
  {Thompson}}, \bibinfo {author} {\bibfnamefont {H.~M.}\ \bibnamefont
  {Aktulga}}, \bibinfo {author} {\bibfnamefont {R.}~\bibnamefont {Berger}},
  \bibinfo {author} {\bibfnamefont {D.~S.}\ \bibnamefont {Bolintineanu}},
  \bibinfo {author} {\bibfnamefont {W.~M.}\ \bibnamefont {Brown}}, \bibinfo
  {author} {\bibfnamefont {P.~S.}\ \bibnamefont {Crozier}}, \bibinfo {author}
  {\bibfnamefont {P.~J.}\ \bibnamefont {in~'t Veld}}, \bibinfo {author}
  {\bibfnamefont {A.}~\bibnamefont {Kohlmeyer}}, \bibinfo {author}
  {\bibfnamefont {S.~G.}\ \bibnamefont {Moore}}, \bibinfo {author}
  {\bibfnamefont {T.~D.}\ \bibnamefont {Nguyen}}, \bibinfo {author}
  {\bibfnamefont {R.}~\bibnamefont {Shan}}, \bibinfo {author} {\bibfnamefont
  {M.~J.}\ \bibnamefont {Stevens}}, \bibinfo {author} {\bibfnamefont
  {J.}~\bibnamefont {Tranchida}}, \bibinfo {author} {\bibfnamefont
  {C.}~\bibnamefont {Trott}}, \ and\ \bibinfo {author} {\bibfnamefont {S.~J.}\
  \bibnamefont {Plimpton}},\ }\bibfield  {title} {\enquote {\bibinfo {title}
  {{LAMMPS} - a flexible simulation tool for particle-based materials modeling
  at the atomic, meso, and continuum scales},}\ }\href {\doibase
  10.1016/j.cpc.2021.108171} {\bibfield  {journal} {\bibinfo  {journal} {Comp.
  Phys. Comm.}\ }\textbf {\bibinfo {volume} {271}},\ \bibinfo {pages} {108171}
  (\bibinfo {year} {2022})}\BibitemShut {NoStop}%
\bibitem [{\citenamefont {Chenoweth}, \citenamefont {van Duin},\ and\
  \citenamefont {Goddard}(2008)}]{Chenoweth_08}%
  \BibitemOpen
  \bibfield  {author} {\bibinfo {author} {\bibfnamefont {K.}~\bibnamefont
  {Chenoweth}}, \bibinfo {author} {\bibfnamefont {A.~C.}\ \bibnamefont {van
  Duin}}, \ and\ \bibinfo {author} {\bibfnamefont {W.~A.}\ \bibnamefont
  {Goddard}},\ }\bibfield  {title} {\enquote {\bibinfo {title} {Reaxff reactive
  force field for molecular dynamics simulations of hydrocarbon oxidation},}\
  }\href {\doibase 10.1021/jp709896w} {\bibfield  {journal} {\bibinfo
  {journal} {The journal of physical chemistry. A}\ } (\bibinfo {year}
  {2008}),\ 10.1021/jp709896w}\BibitemShut {NoStop}%
\bibitem [{\citenamefont {Hoover}(1985)}]{Thermostat}%
  \BibitemOpen
  \bibfield  {author} {\bibinfo {author} {\bibfnamefont {W.~G.}\ \bibnamefont
  {Hoover}},\ }\bibfield  {title} {\enquote {\bibinfo {title} {Canonical
  dynamics: Equilibrium phase-space distributions},}\ }\href {\doibase
  10.1103/PhysRevA.31.1695} {\bibfield  {journal} {\bibinfo  {journal} {Phys.
  Rev. A}\ }\textbf {\bibinfo {volume} {31}},\ \bibinfo {pages} {1695--1697}
  (\bibinfo {year} {1985})}\BibitemShut {NoStop}%
\bibitem [{\citenamefont {Tenenbaum}, \citenamefont {De~Silva},\ and\
  \citenamefont {Langford}(2000)}]{isomap}%
  \BibitemOpen
  \bibfield  {author} {\bibinfo {author} {\bibfnamefont {J.~B.}\ \bibnamefont
  {Tenenbaum}}, \bibinfo {author} {\bibfnamefont {V.}~\bibnamefont {De~Silva}},
  \ and\ \bibinfo {author} {\bibfnamefont {J.~C.}\ \bibnamefont {Langford}},\
  }\bibfield  {title} {\enquote {\bibinfo {title} {A global geometric framework
  for nonlinear dimensionality reduction},}\ }\href {\doibase
  10.1126/science.290.5500.2319} {\bibfield  {journal} {\bibinfo  {journal}
  {Science}\ }\textbf {\bibinfo {volume} {290}},\ \bibinfo {pages} {2319--2323}
  (\bibinfo {year} {2000})}\BibitemShut {NoStop}%
\bibitem [{Note1()}]{Note1}%
  \BibitemOpen
  \bibinfo {note} {To make the calculation of these mean values more robust we
  also calculate the standard deviation $\sigma $. We assume that any distance
  that is more than $2\sigma $ from the mean is an outlier, discard it and then
  recalculate the mean without including these outliers}\BibitemShut {NoStop}%
\bibitem [{\citenamefont {Liu}\ \emph {et~al.}(2012{\natexlab{b}})\citenamefont
  {Liu}, \citenamefont {Liu}, \citenamefont {Cheng}, \citenamefont {Li},
  \citenamefont {Wang},\ and\ \citenamefont {Zheng}}]{Liu_12_deformation}%
  \BibitemOpen
  \bibfield  {author} {\bibinfo {author} {\bibfnamefont {Z.}~\bibnamefont
  {Liu}}, \bibinfo {author} {\bibfnamefont {J.~Z.}\ \bibnamefont {Liu}},
  \bibinfo {author} {\bibfnamefont {Y.}~\bibnamefont {Cheng}}, \bibinfo
  {author} {\bibfnamefont {Z.}~\bibnamefont {Li}}, \bibinfo {author}
  {\bibfnamefont {L.}~\bibnamefont {Wang}}, \ and\ \bibinfo {author}
  {\bibfnamefont {Q.}~\bibnamefont {Zheng}},\ }\bibfield  {title} {\enquote
  {\bibinfo {title} {Interlayer binding energy of graphite: A mesoscopic
  determination from deformation},}\ }\href {\doibase
  10.1103/PhysRevB.85.205418} {\bibfield  {journal} {\bibinfo  {journal} {Phys.
  Rev. B}\ }\textbf {\bibinfo {volume} {85}},\ \bibinfo {pages} {205418}
  (\bibinfo {year} {2012}{\natexlab{b}})}\BibitemShut {NoStop}%
\end{thebibliography}%

\end{document}